\documentclass[11pt]{article}
\usepackage{cite}
\usepackage{amssymb}
\usepackage{amsfonts}
\usepackage{mathrsfs}
\usepackage{graphicx}
\usepackage{amsmath}
\usepackage{amsthm}
\usepackage{epstopdf}
\usepackage{booktabs}
\usepackage{pifont,bm}
\usepackage{tikz}
\usepackage[title]{appendix}
\usepackage{color}
\usepackage{enumerate}
\usepackage{threeparttable}
\usepackage{diagbox}

\theoremstyle{definition}

\makeatletter
\def\@biblabel#1{[#1]}
\makeatother

\makeatletter \@addtoreset{equation}{section}

\begin{document}

\begin{titlepage}
\title{\bf{Long-time asymptotics for the integrable nonlocal  Lakshmanan-Porsezian-Daniel equation with decaying initial value problem
\footnote{
Corresponding authors.\protect\\
\hspace*{3ex} E-mail addresses: ychen@sei.ecnu.edu.cn (Y. Chen)}
}}
\author{Wei-Qi Peng$^{a}$, Yong Chen$^{a,b,*}$\\
\small \emph{$^{a}$School of Mathematical Sciences, Shanghai Key Laboratory of PMMP} \\
\small \emph{East China Normal University, Shanghai, 200241, China} \\
\small \emph{$^{b}$College of Mathematics and Systems Science, Shandong University }\\
\small \emph{of Science and Technology, Qingdao, 266590, China} \\
\date{}}
\thispagestyle{empty}
\end{titlepage}
\maketitle

\vspace{-0.5cm}
\begin{center}
\rule{15cm}{1pt}\vspace{0.3cm}

\parbox{15cm}{\small
{\bf Abstract}\\
\hspace{0.5cm}  In this work, we study the Cauchy problem of integrable nonlocal Lakshmanan-Porsezian-Daniel equation with rapid attenuation of initial data. The basis Riemann-Hilbert problem of integrable nonlocal Lakshmanan-Porsezian-Daniel equation is constructed from Lax pair. Using Deift-Zhou nonlinear steepest descent method, the explicit long-time asymptotic formula of integrable nonlocal Lakshmanan-Porsezian-Daniel equation is derived. For the integrable nonlocal Lakshmanan-Porsezian-Daniel equation, the asymptotic behavior is different from the local model, due to they have different symmetry for the scattering matrix. Besides, since the increase of real stationary phase points also makes the asymptotic behavior have more complex error term which has nine possibilities in our analysis.
}

\vspace{0.5cm}
\parbox{15cm}{\small{

\vspace{0.3cm} \emph{Key words:} Riemann-Hilbert problem; Integrable nonlocal Lakshmanan-Porsezian-Daniel equation; Long-time asymptotics.\\

 } }
\end{center}
\vspace{0.3cm} \rule{15cm}{1pt} \vspace{0.2cm}

\section{Introduction}
In 2013, a $PT$ symmetric nonlocal integrable nonlinear Schr\"{o}dinger (NLS) equation was introduced by Ablowitz and Musslimani \cite{Yang15}. As a new reduction of the Ablowitz-Kaup-Newell-Segur hierarchy, the $PT$ symmetric nonlocal integrable NLS equation is provided with the Lax pair and an infinite number of conservation laws. Compared with the classical NLS equation, the nonlocal NLS equation has a nonlocal term which does not exist in classical one.  After that, more and more scholars pay attention to the
other nonlocal integrable equations and their related properties are studied extensively  \cite{Yang16,Yang17,Yang18,Yang19,Yang20,Yang21,Yang22,Yang23,Yang24,Yang25}.  In this work,
we are committed to the long-time asymptotic behavior of the nonlocal Lakshmanan-Porsezian-
Daniel (LPD) equation  taking the following form \cite{He-CTP}
\begin{align}\label{1}
q_{t}+\frac{1}{2}iq_{xx}(x,t)-iq^{2}(x,t)q^{\ast}(-x, t)-\delta H[q(x,t)]=0,\ (x,t)\in \mathbb{R}\times (0,+\infty),
\end{align}
with
\begin{align}\label{2}
&H[q(x, t)]=-iq_{xxxx}(x,t)+6iq^{\ast}(-x,t)q^{2}_{x}(x,t)+4iq(x,t)q^{\ast}_{x}(-x,t)q_{x}(x,t)\notag\\
&+8iq^{\ast}(-x,t)q(x,t)+q_{xx}(x,t)+2iq^{2}(x,t)q^{\ast}_{xx}(-x,t)-6i(q^{\ast}(-x,t))^{2}q^{3}(x,t),
\end{align}
where $\delta$ is arbitrary positive real parameter. The subscripts represent partial differentiations, and the symbol ``$\ast$'' means the complex conjugation. $q(x,t)$ is the complex function with scaled spatial coordinate $x$ and temporal coordinate $t$, and $q^{\ast}(-x,t)$ is the nonlocal term. A self-induced potential $V(x, t)=q(x, t)q^{\ast}(-x,t)$
admits the $PT$ symmetric restriction $V (x, t)=V^{\ast}(-x, t)$.  The initial data is given by
$q(x, 0)=q_{0}(x)$ which  belongs to  the Schwartz space. The classical LPD equation was first proposed by Lakshmanan, Porsezian, and Daniel through studying the integrable properties of a classical 1-dimensional isotropic
biquadratic Heisenberg spin chain in its continuum limit \cite{He-CTP6,He-CTP7,He-CTP8}.
The LPD equation possesses clearer nonlinear effect due to it has  higher order nonlinear terms than the
NLS equation, including the fourth-order dispersion, the cubic and quintic nonlinearities.
The integrability
of nonlocal LPD equation is demonstrated and its rational soliton solutions were derived by using the degenerate Darboux
transformation in Ref.\cite{He-CTP}. Solitons, periodic waves and modulation instability for the nonlocal LPD equation were studied through the binary Darboux transformation \cite{Wave}. Concise nonsingular solution of the nonlocal LPD equation are given in Matrix
form by the Darboux transformation method \cite{YangAML}. By using inverse scattering transform (IST) method, Xun and Tian obtained the time-periodic pure
soliton solutions of the integrable nonlocal LPD equation, and its integrability as an infinite dimensional Hamilton dynamic system
was established \cite{XunTian}.

In the area of integrable systems, IST is an extremely important technical tool to deal with the nonlinear evolution equation. It was applied first by Gardner, Greene, Kruskal and Miura to solve the Cauchy problem of KdV equation \cite{Peng-Tian9}. The solution of the original nonlinear problem can be obtained by solving a series of linear problems.
Different with the original IST method which relies on the Gel'fand-Levitan-Marchenko (GLM) integral equation, the modern version of IST can be characterized in terms of an associated Riemann-Hilbert(RH) problem
whose solution can be used to represent the solution of the original initial value problem \cite{Peng-Tian10}. Furthermore, the RH problem was also developed to analyse the asymptotic behavior of the solution.  And what's more, Deift and Zhou, inspired by the
early work of Its \cite{Peng-Tian15,Peng-Tian16}, initially proposed the nonlinear steepest descent method to obtain the long-time asymptotic behavior of solutions for the Cauchy problem of  mKdV equation \cite{Peng-Tian14}. The primary idea of Deift-Zhou nonlinear steepest descent method  is to perform a series of deformations on the original oscillation RH problem up to ultimately simplify it into a solvable form. Subsequently, the long-time asymptotic behaviors of various nonlinear integrable models were analysed using the nonlinear steepest descent method \cite{Peng-Tian18,Peng-Tian19,Peng-Tian20,Peng-Tian21,Peng-Tian22,Peng-Tian24,Peng-Tian25,
Peng-Tian26,Peng-Tian27,Peng-Tian28,Peng-Tian29,Peng-Tian30,Peng-Tian31,Peng-Tian32,Peng-Tian33,Peng-Tian34}.

Recently, Dmitry Shepelsky et al. employed the Deift-Zhou method to analyse the long-time behavior of solutions for the Cauchy problem of nonlocal NLS equation \cite{JMP-S}. Following that, the long-time asymptotic behaviors of nonlocal integrable NLS solutions with nonzero boundary conditions and step-like initial data were also investigated by them \cite{Peng16-21,Peng16-22}. Besides, this method was used to discuss the long-time asymptotics for the solution of the nonlocal mKdV equation \cite{Peng16-23} and nonlocal Hirota equation \cite{Peng-Chen}.
As we know, long-time asymptotics for the nonlocal
LPD equation \eqref{1} has not been reported. In this paper,
we focus on the long-time asymptotic behavior of the nonlocal LPD equation \eqref{1}
with the initial data $q_{0}(x)$ rapidly decaying to $0$ as $\mid x\mid\rightarrow \infty$. Compared to the nonlocal NLS equation, nonlocal mKdV equation, and nonlocal Hirota equation, the nonlocal LPD equation has more real stationary
phase points, which leads to a more complicated asymptotics analysis. On the other hand, for the nonlocal LPD equation, the symmetries of its Lax pair are different with the classical LPD equation, then the scattering data for nonlocal LPD equation
satisfy the different symmetry relations.  Another critical difference is that the $\delta_{\lambda_{l}}^{0}$ contains an increasing  $t^{\frac{\pm Im\vartheta(\lambda_{l})}{2}}$ because of $\vartheta(\lambda_{l})(l=1, 2, 3)$ are imaginary in nonlocal case. Finally, the asymptotic behavior for nonlocal LPD equation will behave differently with error term having multiple possibilities.  The pivotal result of this paper is generalized in what follows:

\noindent \textbf{Theorem 1.1.} \emph{
Suppose that  $q(x, t)$ be the solution of the Cauchy problem of the nonlocal LPD equation \eqref{1} with $q_{0}(x)$  lying in the Schwartz space. Assume that the scattering data associated with $q_{0}(x)$ in \eqref{9.2} satisfy\\
(i)$s_{11}(\lambda)$ and $s_{22}(\lambda)$ have no zeros in $\{ \lambda\in \mathbb{C}\mid \mbox{Im} \lambda \leq 0 \}$ and $\{ \lambda\in \mathbb{C}\mid \mbox{Im} \lambda \geq 0 \}$, respectively.\\
(ii)$\int_{-\infty}^{\zeta}\mathrm{d}\ \mbox{arg} (1-r_{1}(s)r_{2}(s))\in (-\frac{\pi}{3}, \frac{\pi}{3})$ for all $\zeta \in \mathbb{R} (i.e. |\mbox{Im} \vartheta(\lambda)|<\frac{1}{6})$, where $r_{1}(\lambda)=\frac{s_{12}^{\ast}(-\lambda^{\ast})}{s_{22}(\lambda)}, r_{2}(\lambda)=\frac{s_{12}(\lambda)}{s_{11}(\lambda)}$.\\
Then, for any positive constants $\delta>0, \epsilon>0$, the  long-time asymptotics  of the solution $q(x, t)$ is
\begin{gather}
q(x, t)=\sum_{l=1}^{3}t^{-\frac{1}{2}+(-1)^{l}\mbox{Im}\vartheta(\lambda_{l})}P_{l}e^{48\delta i\lambda_{l}^{4}t-2i\lambda_{l}^{2}t-(-1)^{l}i\mbox{Re}\vartheta(\lambda_{l})\ln t}+R(\xi, t),\notag\\ t\rightarrow\infty, \xi=\frac{x}{t}\in (-\sqrt{\frac{1}{27\delta}}+\epsilon, \sqrt{\frac{1}{27\delta}}-\epsilon)
\end{gather}
where
\begin{gather}
P_{1}=-\frac{2\sqrt{2\pi}e^{(i\vartheta(\lambda_{1})-\frac{1}{2})\ln (192\delta\lambda_{1}^{2}-4)-2i\vartheta(\lambda_{3})\ln(\lambda_{1}-\lambda_{3})+2i\vartheta(\lambda_{1})\ln(\lambda_{2}-\lambda_{1})+2\chi_{1}(\lambda_{1})
+\frac{\pi}{2}\vartheta(\lambda_{1})-\frac{3}{4}\pi i}}{r_{1}(\lambda_{1})\Gamma(i\vartheta(\lambda_{1}))},\notag\\
P_{2}=\frac{2\sqrt{2\pi}e^{-(i\vartheta(\lambda_{2})+\frac{1}{2})\ln (4-192\delta\lambda_{2}^{2})-2i\vartheta(\lambda_{1})\ln(\lambda_{2}-\lambda_{1})-2i\vartheta(\lambda_{2})\ln(\lambda_{2}-\lambda_{3})+2\chi_{2}(\lambda_{2})
+\frac{\pi}{2}\vartheta(\lambda_{2})-\frac{1}{4}\pi i}}{r_{1}(\lambda_{2})\Gamma(-i\vartheta(\lambda_{2}))},\notag\\
P_{3}=-\frac{2\sqrt{2\pi}e^{(i\vartheta(\lambda_{3})-\frac{1}{2})\ln (192\delta\lambda_{3}^{2}-4)-2i\vartheta(\lambda_{1})\ln(\lambda_{3}-\lambda_{1})+2i\vartheta(\lambda_{3})\ln(\lambda_{2}-\lambda_{3})+2\chi_{3}(\lambda_{3})
+\frac{\pi}{2}\vartheta(\lambda_{3})-\frac{3}{4}\pi i}}{r_{1}(\lambda_{3})\Gamma(i\vartheta(\lambda_{3}))},
\end{gather}
where $\Gamma$ is  Euler's  Gamma function, and $\chi_{l}(\lambda_{l}), \vartheta(\lambda_{l})$ are given in \eqref{21}, \eqref{22}, respectively. The error estimation $R(\xi,t)$ is
\begin{align}
R(\xi,t)=\left\{
\begin{array}{lr}
O(t^{-1+2max\{|\mbox{Im}\vartheta(\lambda_{1})|,|\mbox{Im}\vartheta(\lambda_{2})|,|\mbox{Im}\vartheta(\lambda_{3})|\}}),  \quad (-1)^{l}\mbox{Im}\vartheta(\lambda_{l})> 0,\\
O(t^{-1+2max\{|\mbox{Im}\vartheta(\lambda_{1})|,|\mbox{Im}\vartheta(\lambda_{2})|\}}),  \quad \mbox{Im}\vartheta(\lambda_{1})<0,\mbox{Im}\vartheta(\lambda_{2})>0,\mbox{Im}\vartheta(\lambda_{3})\geq 0,\\
O(t^{-1+2max\{|\mbox{Im}\vartheta(\lambda_{2})|,|\mbox{Im}\vartheta(\lambda_{3})|\}}),  \quad \mbox{Im}\vartheta(\lambda_{1})\geq 0,\mbox{Im}\vartheta(\lambda_{2})>0,\mbox{Im}\vartheta(\lambda_{3})<0,\\
O(t^{-1+2max\{|\mbox{Im}\vartheta(\lambda_{1})|,|\mbox{Im}\vartheta(\lambda_{3})|\}}),  \quad \mbox{Im}\vartheta(\lambda_{1})<0,\mbox{Im}\vartheta(\lambda_{2})\leq 0,\mbox{Im}\vartheta(\lambda_{3})< 0,\\
O(t^{-1+2|\mbox{Im}\vartheta(\lambda_{1})|}),  \quad\qquad\qquad\qquad \mbox{Im}\vartheta(\lambda_{1})<0,\mbox{Im}\vartheta(\lambda_{2})\leq 0,\mbox{Im}\vartheta(\lambda_{3})\geq 0,\\
O(t^{-1+2|\mbox{Im}\vartheta(\lambda_{2})|}),  \quad\qquad\qquad\qquad \mbox{Im}\vartheta(\lambda_{1})\geq 0,\mbox{Im}\vartheta(\lambda_{2})> 0,\mbox{Im}\vartheta(\lambda_{3})\geq 0,\\
O(t^{-1+2|\mbox{Im}\vartheta(\lambda_{3})|}),  \quad\qquad\qquad\qquad \mbox{Im}\vartheta(\lambda_{1})\geq 0,\mbox{Im}\vartheta(\lambda_{2})\leq 0,\mbox{Im}\vartheta(\lambda_{3})< 0,\\
O(t^{-1}\ln t),  \qquad\qquad\quad \mbox{Im}\vartheta(\lambda_{l})=0, (-1)^{s}\mbox{Im}\vartheta(\lambda_{s})\leq 0, s=1,2,3\ \mbox{and}\ s \neq l, \\
O(t^{-1}),  \qquad\qquad\qquad\qquad\qquad\qquad\qquad\qquad (-1)^{l}\mbox{Im}\vartheta(\lambda_{l})< 0.\\
  \end{array}
\right.
\end{align}
}

\textbf{Organization of the paper:} In section 2, the fundamental RH problem is constructed by the direct scattering analysis and the solution of  nonlocal LPD equation is expressed by the fundamental RH problem.  In section 3, through  the phase analysis and a series of deformations, we derive a model RH problem. Then, the long-time asymptotics of the solution for the nonlocal LPD equation is presented via solving the model RH problem.

\section{Inverse scattering transform and the RH problem}
In this section, we aim to construct the fundamental RH problem through the direct scattering analysis.  The nonlocal LPD equation admits the following spectral problem
\begin{gather}
\Psi_{x}=L\Psi,\qquad L\equiv \lambda J+U,\notag\\
\Psi_{t}=M\Psi,\qquad M\equiv \lambda^{2}J+\lambda U+\frac{1}{2}V+\delta V_{1},\label{3}
\end{gather}
with
\begin{gather}
J=\left(\begin{array}{cc}
    i  &  0\\
    0  &  -i\\
\end{array}\right),\quad U=\left(\begin{array}{cc}
    0  &  q(x,t)\\
    q^{\ast}(-x,t) &  0\\
\end{array}\right),\notag\\
V=\left(\begin{array}{cc}
    iq(x,t)q^{\ast}(-x,t)  &  -iq_{x}(x,t)\\
    iq^{\ast}_{x}(-x,t) &  -iq(x,t)q^{\ast}(-x,t)\\
\end{array}\right),\quad V_{1}=\left(\begin{array}{cc}
    iA(x,t)  &  B(x,t)\\
    -C(x,t) &  -iA(x,t)\\
\end{array}\right), \notag\\
A(x,t)=-8\lambda^{4}-4q^{\ast}(-x,t)q(x,t)\lambda^{2}-2i(q(x,t)q_{x}^{\ast}(-x,t)-q^{\ast}(-x,t)q_{x}(x,t))\lambda\notag\\
-3q^{2}(x,t)(q^{\ast}(-x,t))^{2}
-q_{x}^{\ast}(-x,t)q_{x}(x,t)+q(x,t)q_{xx}^{\ast}(-x,t)+q^{\ast}(-x,t)q_{xx}(x,t),\notag\\
B(x,t)=-8q(x,t)\lambda^{3}+4iq_{x}(x,t)\lambda^{2}+2q_{xx}(x,t)\lambda\notag\\
-4q^{\ast}(-x,t)q^{2}(x,t)\lambda-iq_{xxx}(x,t)+6iq(x,t)q^{\ast}(-x,t)q_{x}(x,t),\notag\\
C(x,t)=8q^{\ast}(-x,t)\lambda^{3}+4iq^{\ast}_{x}(-x,t)\lambda^{2}-2q^{\ast}_{xx}(-x,t)\lambda\notag\\
+4(q^{\ast}(-x,t))^{2}q(x,t)\lambda-iq^{\ast}_{xxx}(-x,t)+6iq^{\ast}(-x,t)q(x,t)q^{\ast}_{x}(-x,t),
\label{4}
\end{gather}
where $\lambda$ is the spectral parameter, $\Psi=\Psi\left(x, t, \lambda\right)$ denotes the eigenfunction.
Defining the Jost solutions $\Psi_{\pm}=\mu_{\pm}e^{i[\lambda x+(\lambda^{2}-8\delta \lambda^{4})t]\sigma_{3}}$, of which $\mu_{\pm}$ are the solution of following Volterra integral equations
\begin{align}\label{5}
\mu_{-}=I+\int_{-\infty}^{x}e^{i\lambda (x-x')\hat{\sigma}_{3}}[U(x',t)\mu_{1}(x',t, \lambda)]\mathrm{d}x',\notag\\
\mu_{+}=I-\int_{x}^{+\infty}e^{i\lambda (x-x')\hat{\sigma}_{3}}[U(x',t)\mu_{2}(x',t, \lambda)]\mathrm{d}x'.
\end{align}
Suppose $q\in L^{1} (\mathbb{R}^{\pm})$, then $\mu_{\pm}(x,t, \lambda)$ have the following properties:
$\mu_{+1}(x, t, \lambda)$ and $\mu_{-2}(x, t, \lambda)$ are analytical and bounded in $\{ \lambda\in \mathbb{C}\mid \mbox{Im} \lambda > 0 \}$; $\mu_{-1}(x, t, \lambda)$ and $\mu_{+2}(x, t, \lambda)$  are analytical and bounded  in $\{ \lambda\in \mathbb{C}\mid \mbox{Im} \lambda < 0 \}$;  $\mu_{\pm}(x,t, \lambda)\rightarrow I$ as $\lambda\rightarrow \infty$; There also are $\det \mu_{\pm}(x, t, \lambda)= 1$ for all $x, t$ and $k$.

As the simultaneous solutions of spectral problem \eqref{3}, $\Psi_{\pm}$ satisfy the following linear relation via
defining a scattering matrix $S(\lambda), \lambda \in \mathbb{R}$, given by
\begin{align}\label{6}
\Psi_{-}(x, t, \lambda)=\Psi_{+}(x, t, \lambda)S(\lambda),\qquad \lambda\in \mathbb{R}.
\end{align}
Furthermore, according to the symmetry
\begin{align}\label{7}
\Psi_{\mp}(x, t, \lambda)=\sigma_{2}\Psi_{\pm}^{\ast}(-x, t, -\lambda^{\ast})
\end{align}
and combine the above expression, the $S(\lambda)$ can be written in the
form
\begin{align}\label{8}
S(\lambda)=\left(\begin{array}{cc}
    s_{11}(\lambda)  &  s_{12}(\lambda)\\
    s_{12}^{\ast}(-\lambda^{\ast}) &  s_{22}(\lambda)\\
\end{array}\right), \quad \lambda\in \mathbb{R},
\end{align}
where the scattering data $s_{11}(\lambda), s_{22}(\lambda)$ arrive at the following symmetry relations
\begin{align}\label{9}
s_{11}(\lambda)=s_{11}^{\ast}(-\lambda^{\ast}),\quad s_{22}(\lambda)=s_{22}^{\ast}(-\lambda^{\ast}),
\end{align}
which implies that the scattering data $s_{11}(\lambda)$ and $s_{22}(\lambda)$
for the nonlocal LPD equation satisfy the different symmetry relations compared with the case of the local LPD equation whose symmetry relations are $s_{11}(\lambda)=s_{22}^{\ast}(\lambda^{\ast})$.

Naturally,  through $S(\lambda)=(\Psi_{+}(x,0, \lambda))^{-1}\Psi_{-}(x,0, \lambda)=e^{-i\lambda x\sigma_{3}}\mu_{+}^{-1}(x,0, \lambda)\mu_{-}(x,0, \lambda)e^{i\lambda x\sigma_{3}},$ we can uniquely give the scattering matrix $S(\lambda)$ using the initial data $q(x, 0)$. The specific form is as follows
\begin{gather}
s_{11}(\lambda)=\lim_{x\rightarrow +\infty} [\mu_{-}]_{11}(x, \lambda),\notag\\
s_{12}(\lambda)=\lim_{x\rightarrow +\infty} e^{-2i\lambda x}[\mu_{-}]_{12}(x, \lambda),\notag\\
s_{22}(\lambda)=\lim_{x\rightarrow +\infty} [\mu_{-}]_{22}(x, \lambda).\label{9.2}
\end{gather}
where
\begin{gather}
[\mu_{-}]_{11}(x,\lambda)=1+\int_{-\infty}^{x}q_{0}(x')[\mu_{-}]_{21}(x',\lambda)\mathrm{d}x',\notag\\
[\mu_{-}]_{12}(x,\lambda)=\int_{-\infty}^{x}e^{2i\lambda(x-x')}q_{0}(x')[\mu_{-}]_{22}(x',\lambda)\mathrm{d}x',\notag\\
[\mu_{-}]_{21}(x,\lambda)=\int_{-\infty}^{x}e^{-2i\lambda(x-x')}q^{\ast}_{0}(-x')[\mu_{-}]_{11}(x',\lambda)\mathrm{d}x',\notag\\
[\mu_{-}]_{22}(x,\lambda)=1+\int_{-\infty}^{x}q^{\ast}_{0}(-x')[\mu_{-}]_{12}(x',\lambda)\mathrm{d}x'.\label{9.1}
\end{gather}

Similarly, according to the analyticity of $\mu_{\pm}$, we know that $s_{11}(\lambda)$ is analytic in the half-plane $\{ \lambda\in \mathbb{C}\mid \mbox{Im} \lambda < 0 \}$ and continuous
in $\{ \lambda\in \mathbb{C}\mid \mbox{Im} \lambda \leq 0 \}$, $s_{22}(\lambda)$ is analytic in the half-plane $\{ \lambda\in \mathbb{C}\mid \mbox{Im} \lambda > 0 \}$ and continuous
in $\{ \lambda\in \mathbb{C}\mid \mbox{Im} \lambda \geq 0 \}$, and $S(\lambda)\rightarrow I$ \mbox{as}  $\lambda\rightarrow \infty$. Furthermore, $\det S(\lambda)=1$ for $\lambda\in \mathbb{R}$.

In this paper, we suppose that $s_{11}(\lambda)$ and $s_{22}(\lambda)$ have no zeros in $\{ \lambda\in \mathbb{C}\mid \mbox{Im} \lambda \leq 0 \}$ and $\{ \lambda\in \mathbb{C}\mid \mbox{Im} \lambda \geq 0 \}$, respectively. Then, we can construct a fundamental RH problem by defining the matrix-valued function $M$ as follows:
\begin{align}\label{10}
M_{+}(x, t, \lambda)=(\frac{\mu_{+1}}{s_{22}},\mu_{-2}),\qquad M_{-}(x, t, \lambda)=(\mu_{-1},\frac{\mu_{+2}}{s_{11}}),
\end{align}
where $\pm$ stand for analyticity in $\{ \lambda\in \mathbb{C}\mid \mbox{Im} \lambda > 0 \}$ and $\{ \lambda\in \mathbb{C}\mid \mbox{Im} \lambda < 0 \}$, respectively.

\noindent \textbf{Riemann-Hilbert Problem}  \emph{
$M(x, t, \lambda)$ satisfies the following RH problem:
\begin{align}\label{11}
\left\{
\begin{array}{lr}
M(x, t, \lambda)\ \mbox{is analytic in} \ \mathbb{C }\setminus \mathbb{R},\\
M_{+}(x, t, \lambda)=M_{-}(x, t, \lambda)J(x, t, \lambda), \qquad \lambda\in \mathbb{R},\\
M(x, t, \lambda)\rightarrow I,\qquad \lambda\rightarrow \infty,\\
  \end{array}
\right.
\end{align}
with the jump matrix $J(x, t, \lambda)$ being
\begin{align}\label{12}
J(x, t, \lambda)=\left(\begin{array}{cc}
    1-r_{1}(\lambda)r_{2}(\lambda)  &  r_{2}(\lambda)e^{2i\theta(x, t, \lambda)}\\
  -r_{1}(\lambda)e^{-2i\theta(x, t, \lambda)} &  1\\
\end{array}\right),
\end{align}
where $r_{1}(\lambda)=\frac{s_{12}^{\ast}(-\lambda^{\ast})}{s_{22}(\lambda)}, r_{2}(\lambda)=\frac{s_{12}(\lambda)}{s_{11}(\lambda)}, \theta(x, t, \lambda)=\lambda x+(\lambda^{2}-8\delta \lambda^{4})t$.
}

Let
\begin{align}\label{13}
M(x, t, \lambda)=I+\frac{1}{\lambda}M_{1}(x, t)+O(\frac{1}{\lambda^{2}}),\qquad \lambda\rightarrow \infty,
\end{align}
then the solution $q(x, t)$ of the nonlocal LPD equation \eqref{1} is expressed
by the solutions of fundamental RH problem
\begin{align}\label{14}
q(x, t)=-2i\left[M_{1}(x, t, \lambda)\right]_{12}=-2i\lim_{\lambda\rightarrow\infty}\lambda \left[M(x, t, \lambda)\right]_{12}.
\end{align}

\section{The long-time behavior for the nonlocal LPD equation}
\begin{figure}[!htb]
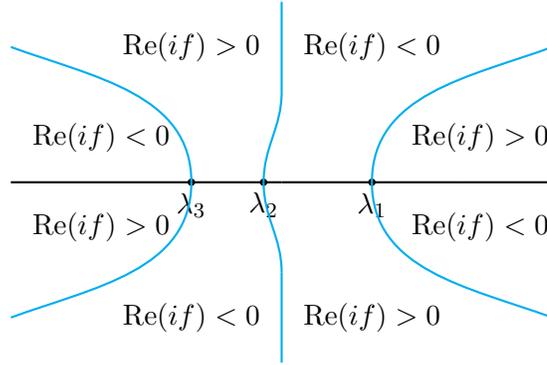

\centering
\caption{\footnotesize The signature table for $\mbox{Re}( if)$ in the complex $\lambda$-plane.}
\label{F1}
\end{figure}

In this section, we aim to transform the associated original RH problem \eqref{11} to
a solvable RH problem and then find the explicitly asymptotic formula for the nonlocal LPD
equation \eqref{1}. Let $\xi=\frac{x}{t}$, $f(\xi, \lambda)$ can be defined by
\begin{align}\label{15}
f(\xi, \lambda)=\lambda\xi+\lambda^{2}-8\delta\lambda^{4}.
\end{align}
Then, we take
\begin{align}\label{16}
-\sqrt{\frac{1}{27\delta}}<\xi<\sqrt{\frac{1}{27\delta}},
\end{align}
it follows that there are three different real solutions ( stationary points) for $\frac{\mathrm{d}f}{\mathrm{d}\lambda}=0$, given by
\begin{align}\label{17}
\lambda_{1}=\sqrt[3]{\frac{\xi}{64\delta}+\sqrt{(\frac{\xi}{64\delta})^{2}-(\frac{1}{48\delta})^{3}}}+\sqrt[3]{\frac{\xi}{64\delta}
-\sqrt{(\frac{\xi}{64\delta})^{2}-(\frac{1}{48\delta})^{3}}}, \notag\\ \lambda_{2}=\omega\sqrt[3]{\frac{\xi}{64\delta}+\sqrt{(\frac{\xi}{64\delta})^{2}-(\frac{1}{48\delta})^{3}}}+\omega^{2}\sqrt[3]{\frac{\xi}{64\delta}
-\sqrt{(\frac{\xi}{64\delta})^{2}-(\frac{1}{48\delta})^{3}}}, \notag\\
\lambda_{3}=\omega^{2}\sqrt[3]{\frac{\xi}{64\delta}+\sqrt{(\frac{\xi}{64\delta})^{2}-(\frac{1}{48\delta})^{3}}}+\omega\sqrt[3]{\frac{\xi}{64\delta}
-\sqrt{(\frac{\xi}{64\delta})^{2}-(\frac{1}{48\delta})^{3}}},
\end{align}
where $\omega=\frac{-1+\sqrt{3}i}{2}$. In this situation, the signature distribution for $\mbox{Re} (i f)$ is shown in Figure. \ref{F1}.  The following analysis of this paper restricts $\xi$ to region $\xi\in (-\sqrt{\frac{1}{27\delta}}+\epsilon, \sqrt{\frac{1}{27\delta}}-\epsilon)$ for any positive constant $\epsilon$.\\
\centerline{
\begin{tikzpicture}[scale=1.2]
\draw[-][thick](-3,0)--(-2,0);
\draw[-][thick](-2.0,0)--(-1.0,0)node[below]{$\lambda_{3}$};
\draw[-][thick](-1,0)--(0,0);
\draw[-][thick](0,0)--(1,0)node[below]{$\lambda_{1}$};
\draw[-][thick](1,0)--(2,0);
\draw[fill] (1,0) circle [radius=0.035];
\draw[fill] (-1,0) circle [radius=0.035];
\draw[fill] (-0.2,0) circle [radius=0.035]node[below]{$\lambda_{2}$};
\draw[-][thick](2.0,0)--(3.0,0);
\draw [-,thick, cyan] (1,0) to [out=90,in=-160] (3,1.5);
\draw [-,thick, cyan] (1,0) to [out=-90,in=160] (3,-1.5);
\draw [-,thick, cyan] (-1,0) to [out=90,in=-20] (-3,1.5);
\draw [-,thick, cyan] (-1,0) to [out=-90,in=20] (-3,-1.5);
\draw [-,thick, cyan] (-0.2,0) to [out=90,in=-90] (0,1);
\draw [-,thick, cyan] (-0.2,0) to [out=-90,in=90] (0,-1);
\draw[-][thick, cyan](0,1)--(0,2);
\draw[-][thick, cyan](0,-1)--(0,-2);
\draw[fill] (-1,1.5) node{$\mbox{Re}(if)>0$};
\draw[fill] (-1,-1.5) node{$\mbox{Re}(if)<0$};
\draw[fill] (-2,0.5) node{$\mbox{Re}(if)<0$};
\draw[fill] (2.2,0.5) node{$\mbox{Re}(if)>0$};
\draw[fill] (2.2,-0.5) node{$\mbox{Re}(if)<0$};
\draw[fill] (-2,-0.5) node{$\mbox{Re}(if)>0$};
\draw[fill] (1,1.5) node{$\mbox{Re}(if)<0$};
\draw[fill] (1,-1.5) node{$\mbox{Re}(if)>0$};
\end{tikzpicture}}\\

\subsection{Factorization of the jump matrix}
The jump matrix $J(x, t, \lambda)$ should be decomposed into
\begin{align}\label{18}
J(x, t, \lambda)=\left\{
\begin{array}{lr}
\left(\begin{array}{cc}
     1 &  r_{2}(\lambda)e^{2ift} \\
     0  &  1\\
\end{array}\right)\left(\begin{array}{cc}
     1 &  0\\
     -r_{1}(\lambda)e^{-2ift}  &  1\\
\end{array}\right),\\
\left(\begin{array}{cc}
     1 &  0\\
     -\frac{r_{1}(\lambda)e^{-2ift}}{1-r_{1}(\lambda)r_{2}(\lambda)}  &  1\\
\end{array}\right)\left(\begin{array}{cc}
     1-r_{1}(\lambda)r_{2}(\lambda) &  0 \\
     0  &  \frac{1}{1-r_{1}(\lambda)r_{2}(\lambda)}\\
\end{array}\right)
\left(\begin{array}{cc}
    1 &  \frac{r_{2}(\lambda)e^{2ift}}{1-r_{1}(\lambda)r_{2}(\lambda)}\\
    0  &  1\\
\end{array}\right).
  \end{array}
\right.
\end{align}
It is necessary to define the following RH problem about the function $\delta(\lambda)$
\begin{align}\label{19}
\left\{
\begin{array}{lr}
\delta_{+}( \lambda)=(1-r_{1}(\lambda)r_{2}(\lambda))\delta_{-}( \lambda), \qquad \lambda\in(\lambda_{3},\lambda_{2})\cup (\lambda_{1},+\infty),\\
\delta(\lambda)\rightarrow 1,\qquad \lambda\rightarrow \infty.
  \end{array}
\right.
\end{align}
Using the Plemelj formula, we get
\begin{align}\label{20}
\delta(\lambda)&=\exp\left\{\frac{1}{2\pi i}(\int_{\lambda_{3}}^{\lambda_{2}}+\int_{\lambda_{1}}^{\infty})\frac{\ln(1-r_{1}(s)r_{2}(s))}{s-\lambda}\mathrm{d}s\right\}
=(\lambda-\lambda_{3})^{-i\vartheta(\lambda_{3})}\left(\frac{\lambda-\lambda_{1}}{\lambda-\lambda_{2}}\right)^{-i\vartheta(\lambda_{1})}e^{\chi_{1}(\lambda)}\notag\\
&=(\lambda-\lambda_{1})^{-i\vartheta(\lambda_{1})}\left(\frac{\lambda-\lambda_{3}}{\lambda-\lambda_{2}}\right)^{-i\vartheta(\lambda_{2})}e^{\chi_{2}(\lambda)}
=(\lambda-\lambda_{1})^{-i\vartheta(\lambda_{1})}\left(\frac{\lambda-\lambda_{3}}{\lambda-\lambda_{2}}\right)^{-i\vartheta(\lambda_{3})}e^{\chi_{3}(\lambda)},
\end{align}
where
\begin{align}\label{21}
\chi_{1}(\lambda)=\frac{1}{2\pi i}\left[\int_{\lambda_{1}}^{\lambda_{2}}\ln\left(\frac{1-r_{1}(s)r_{2}(s)}
{1-r_{1}(\lambda_{1})r_{2}(\lambda_{1})}\right)\frac{\mathrm{d}s}{s-\lambda}-\int_{\lambda_{3}}^{\infty}\ln\left(\lambda-s\right)\mathrm{d}\ln(1-r_{1}(s)r_{2}(s))\right],\notag\\
\chi_{2}(\lambda)=\frac{1}{2\pi i}\left[\int_{\lambda_{3}}^{\lambda_{2}}\ln\left(\frac{1-r_{1}(s)r_{2}(s)}
{1-r_{1}(\lambda_{2})r_{2}(\lambda_{2})}\right)\frac{\mathrm{d}s}{s-\lambda}-\int_{\lambda_{1}}^{\infty}\ln\left(\lambda-s\right)\mathrm{d}\ln(1-r_{1}(s)r_{2}(s))\right],\notag\\
\chi_{3}(\lambda)=\frac{1}{2\pi i}\left[\int_{\lambda_{3}}^{\lambda_{2}}\ln\left(\frac{1-r_{1}(s)r_{2}(s)}
{1-r_{1}(\lambda_{3})r_{2}(\lambda_{3})}\right)\frac{\mathrm{d}s}{s-\lambda}-\int_{\lambda_{1}}^{\infty}\ln\left(\lambda-s\right)\mathrm{d}\ln(1-r_{1}(s)r_{2}(s))\right],
\end{align}
with
\begin{align}\label{22}
\vartheta(\lambda_{1})=-\frac{1}{2\pi}\ln(1-r_{1}(\lambda_{1})r_{2}(\lambda_{1})),\notag\\
\vartheta(\lambda_{2})=-\frac{1}{2\pi}\ln(1-r_{1}(\lambda_{2})r_{2}(\lambda_{2})),\notag\\
\vartheta(\lambda_{3})=-\frac{1}{2\pi}\ln(1-r_{1}(\lambda_{3})r_{2}(\lambda_{3})),
\end{align}
so that
\begin{align}\label{23}
\mbox{Im}\vartheta(\lambda_{l})=-\frac{1}{2\pi}\int_{-\infty}^{\lambda_{l}}\mathrm{d}\ \mbox{arg} (1-r_{1}(s)r_{2}(s)),\ l=1, 2, 3.
\end{align}
Assuming that $\int_{-\infty}^{\lambda_{l}}\mathrm{d}\ \mbox{arg} (1-r_{1}(s)r_{2}(s))\in (-\frac{\pi}{3}, \frac{\pi}{3})$, one has
\begin{align}\label{24}
|\mbox{Im}\vartheta(\lambda)|<\frac{1}{6}, \qquad \lambda\in \mathbb{R},
\end{align}
which indicates that $\ln(1-r_{1}(\lambda)r_{2}(\lambda))$ is single-valued, and the singularity of $\delta(\lambda, \xi)$ at $\lambda=\lambda_{l}$ is square integrable.

\begin{figure}[!htb]
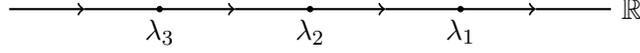

\centering
\caption{\footnotesize The jump contour $\mathbb{R}$.}
\label{F2}
\end{figure}

Let
\begin{align}\label{25}
\tilde{M}(x, t,\lambda)=M(x, t,\lambda)\delta^{-\sigma_{3}}(\lambda),
\end{align}
then $\tilde{M}$ is the solution of  RH problem on the jump contour $\mathbb{R}$ shown in Figure. \ref{F2},
\begin{align}\label{26}
\left\{
\begin{array}{lr}
\tilde{M}_{+}(x, t, \lambda)=\tilde{M}_{-}(x, t, \lambda)\tilde{J}(x, t, \lambda), \qquad \lambda\in\mathbb{R},\\
\tilde{M}(x, t, \lambda)\rightarrow I,\qquad \lambda\rightarrow \infty,
  \end{array}
\right.
\end{align}
where
\begin{align}\label{27}
\tilde{J}(x, t, \lambda)=\left\{
\begin{array}{lr}
\left(\begin{array}{cc}
     1 &  r_{2}(\lambda)\delta^{2}e^{2ift} \\
     0  &  1\\
\end{array}\right)\left(\begin{array}{cc}
     1 &  0\\
     -r_{1}(\lambda)\delta^{-2}e^{-2ift}  &  1\\
\end{array}\right),\quad \lambda\in(\lambda_{2},\lambda_{1})\cup(-\infty, \lambda_{3}),\\
\left(\begin{array}{cc}
     1 &  0\\
     -\frac{r_{1}(\lambda)\delta_{-}^{-2}e^{-2ift}}{1-r_{1}(\lambda)r_{2}(\lambda)}  &  1\\
\end{array}\right)
\left(\begin{array}{cc}
    1 &  \frac{r_{2}(\lambda)\delta_{+}^{2}e^{2ift}}{1-r_{1}(\lambda)r_{2}(\lambda)}\\
    0  &  1\\
\end{array}\right),\quad \lambda\in(\lambda_{3},\lambda_{2})\cup(\lambda_{1}, +\infty).
  \end{array}
\right.
\end{align}
\\
\centerline{\begin{tikzpicture}[scale=1.0]
\draw[->][thick](-3,0)--(-2,0);
\draw[-][thick](-2.0,0)--(-1.0,0)node[below]{$\lambda_{3}$};
\draw[->][thick](-1,0)--(0,0);
\draw[-][thick](0,0)--(1,0)node[below]{$\lambda_{2}$};
\draw[->][thick](1,0)--(2,0);
\draw[fill] (1,0) circle [radius=0.035];
\draw[fill] (-1,0) circle [radius=0.035];
\draw[-][thick](2.0,0)--(3.0,0);
\draw[fill] (3,0) circle [radius=0.035]node[below]{$\lambda_{1}$};
\draw[->][thick](3,0)--(4,0);
\draw[-][thick](4,0)--(5,0)node[right]{$\mathbb{R}$};
\end{tikzpicture}}\\

\subsection{RH problem transformations}
According to the Refs.\cite{Peng-Tian14,JLenells}, we know that the reflection coefficients $r_{j}(\lambda), \frac{r_{j}(\lambda)}{1-r_{1}(\lambda)r_{2}(\lambda)}, j=1, 2$ can be approximated by some rational functions with good error control.
However, if we assume that the initial data $q_{0}(x)$ decay exponentially to zero
as $\mid x \mid \rightarrow \infty$, then $r_{j}(\lambda)$ are analytic in a band containing $\lambda\in \mathbb{R}$, so there is no need to use rational approximation for the next RH problem transformation below. Thus, if desired, the analytical approximations of $r_{j}(\lambda)$ and $\frac{r_{j}(\lambda)}{1-r_{1}(\lambda)r_{2}(\lambda)}$ remain the same notations.
\\
\centerline{\begin{tikzpicture}[scale=1.5]
\draw[->][thick](-3,0)--(-2,0);
\draw[-][thick](-2.0,0)--(-1.0,0)node[below]{$\lambda_{3}$};
\draw[->][thick](-1,0)--(0,0);
\draw[->][thick](0,1)--(0.5,0.5);
\draw[-][thick](0.5,0.5)--(1,0);
\draw[->][thick](0,-1)--(0.5,-0.5);
\draw[-][thick](0.5,-0.5)--(1,0);
\draw[->][thick](2,-1)--(1.5,-0.5);
\draw[-][thick](1.5,-0.5)--(1,0);
\draw[->][thick](2,1)--(1.5,0.5);
\draw[-][thick](1.5,0.5)--(1,0);
\draw[-][thick](0,1)--(-0.5,0.5);
\draw[->][thick](-1,0)--(-0.5,0.5);
\draw[-][thick](0,-1)--(-0.5,-0.5);
\draw[->][thick](-1,0)--(-0.5,-0.5);
\draw[->][thick](-1,0)--(-1.5,0.5);
\draw[-][thick](-1.5,0.5)--(-2,1);
\draw[->][thick](-1,0)--(-1.5,-0.5);
\draw[-][thick](-1.5,-0.5)--(-2,-1);
\draw[-][thick](0,0)--(1,0)node[below]{$\lambda_{2}$};
\draw[->][thick](1,0)--(2,0);
\draw[fill] (1,0) circle [radius=0.035];
\draw[fill] (3,0) circle [radius=0.035]node[below]{$\lambda_{1}$};
\draw[->][thick](3,0)--(4,0);
\draw[-][thick](4,0)--(5,0);
\draw[fill] (-1,0) circle [radius=0.035];
\draw[-][thick](2.0,0)--(3.0,0);
\draw[fill] (1,1.0) node{$\Omega_{5}$};
\draw[fill] (1,-1.0) node{$\Omega_{6}$};
\draw[fill] (0,0.5) node{$\Omega_{2}$};
\draw[fill] (0,-0.5) node{$\Omega_{3}$};
\draw[fill] (2,-0.5) node{$\Omega_{4}$};
\draw[fill] (2,0.5) node{$\Omega_{1}$};
\draw[fill] (-2,-0.5) node{$\Omega_{4}$};
\draw[fill] (-2,0.5) node{$\Omega_{1}$};
\draw[fill] (4,0.5) node{$\Omega_{2}$};
\draw[fill] (4,-0.5) node{$\Omega_{3}$};
\draw[-][thick](2,1)--(2.5,0.5);
\draw[<-][thick](2.5,0.5)--(3,0);
\draw[-][thick](2,-1)--(2.5,-0.5);
\draw[<-][thick](2.5,-0.5)--(3,0);
\draw[->][thick](3,0)--(3.5,0.5);
\draw[-][thick](3.5,0.5)--(4,1);
\draw[->][thick](3,0)--(3.5,-0.5);
\draw[-][thick](3.5,-0.5)--(4,-1);
\draw[fill] (-1.4,0.6) node{$\gamma_{1}$};
\draw[fill] (-1.4,-0.6) node{$\gamma_{4}$};
\draw[fill] (-0.6,0.6) node{$\gamma_{2}$};
\draw[fill] (-0.6,-0.6) node{$\gamma_{3}$};
\draw[fill] (0.6,0.6) node{$\gamma_{2}$};
\draw[fill] (0.6,-0.6) node{$\gamma_{3}$};
\draw[fill] (1.4,0.6) node{$\gamma_{1}$};
\draw[fill] (1.4,-0.6) node{$\gamma_{4}$};
\draw[fill] (2.6,0.6) node{$\gamma_{1}$};
\draw[fill] (2.6,-0.6) node{$\gamma_{4}$};
\draw[fill] (3.4,0.6) node{$\gamma_{2}$};
\draw[fill] (3.4,-0.6) node{$\gamma_{3}$};
\end{tikzpicture}}\\
\begin{figure}[!htb]
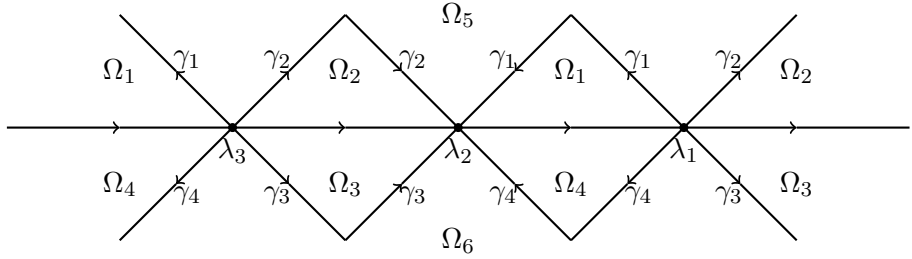

\centering
\caption{\footnotesize The jump contour $\Sigma$  and domains $\Omega_{j}(j=1,\cdots, 6)$.}
\label{F3}
\end{figure}

Next, we perform the first RH problem transformation by defining $\tilde{\tilde{M}}(x, t, \lambda)$ as follows (see Figure \ref{F3}):

\begin{align}\label{28}
\tilde{\tilde{M}}=\left\{
\begin{array}{lr}
\tilde{M}(\lambda), \qquad \qquad \quad \lambda\in \Omega_{5}\cup \Omega_{6},\\
\tilde{M}(\lambda)\left(\begin{array}{cc}
    1 &  0\\
    r_{1}(\lambda)\delta^{-2}e^{-2ift}  &  1\\
\end{array}\right), \qquad \lambda\in \Omega_{1},\\
\tilde{M}(\lambda)\left(\begin{array}{cc}
    1 &  -\frac{r_{2}(\lambda)\delta^{2}}{1-r_{1}(\lambda)r_{2}(\lambda)}e^{2ift}\\
    0  &  1\\
\end{array}\right), \qquad \lambda\in \Omega_{2},\\
\tilde{M}(\lambda)\left(\begin{array}{cc}
    1 &  0\\
    -\frac{r_{1}(\lambda)\delta^{-2}}{1-r_{1}(\lambda)r_{2}(\lambda)}e^{-2ift}  &  1\\
\end{array}\right), \qquad \lambda\in \Omega_{3},\\
\tilde{M}(\lambda)\left(\begin{array}{cc}
    1 &  r_{2}(\lambda)\delta^{2}e^{2ift}\\
    0  &  1\\
\end{array}\right), \qquad \lambda\in \Omega_{4},\\
  \end{array}
\right.
\end{align}
Then,  the following RH problem on the contour $\Sigma$ is obtained
\begin{align}\label{29}
\left\{
\begin{array}{lr}
\tilde{\tilde{M}}_{+}(x, t, \lambda)=\tilde{\tilde{M}}_{-}(x, t, \lambda)\tilde{\tilde{J}}(x, t, \lambda), \qquad \lambda\in\Sigma,\\
\tilde{\tilde{M}}(x, t, \lambda)\rightarrow I,\qquad \lambda\rightarrow \infty,
  \end{array}
\right.
\end{align}
where the jump matrix become
\begin{align}\label{30}
\tilde{\tilde{J}}=\left\{
\begin{array}{lr}
\left(\begin{array}{cc}
    1 &  0\\
    r_{1}(\lambda)\delta^{-2}e^{-2ift}  &  1\\
\end{array}\right), \qquad \qquad \lambda\in \gamma_{1},\\
\left(\begin{array}{cc}
    1 &  \frac{r_{2}(\lambda)\delta^{2}}{1-r_{1}(\lambda)r_{2}(\lambda)}e^{2ift}\\
    0  &  1\\
\end{array}\right),  \qquad\quad\ \lambda\in \gamma_{2},\\
\left(\begin{array}{cc}
    1 &  0\\
    -\frac{r_{1}(\lambda)\delta^{-2}}{1-r_{1}(\lambda)r_{2}(\lambda)}e^{-2ift}  &  1\\
\end{array}\right), \qquad \lambda\in \gamma_{3},\\
\left(\begin{array}{cc}
    1 &  -r_{2}(\lambda)\delta^{2}e^{2ift}\\
    0  &  1\\
\end{array}\right), \qquad \lambda\in \gamma_{4}.
  \end{array}
\right.
\end{align}
According to the signature table for $\mbox{Re}( if)$ (see Figure \ref{F1}),  we easily find that the jump matrix $\tilde{\tilde{J}}(x, t, \lambda)$ converges rapidly the identity matrix as $t\rightarrow \infty$  for all  $\Sigma\setminus\{\lambda_{l}\}$, which means that we only need focus on neighborhood of $\lambda_{l}$ for the long-time asymptotics analysis.

\subsection{Reduction to a model RH problem}
The purpose in this part is to carry out the scaling transformation to separate the time $t$ from
the jump matrix. To realize it,  a scaling transformation is introduced by
\begin{align}\label{31}
z=T_{1}(\lambda)=\sqrt{4t(48\delta\lambda_{1}^{2}-1)}(\lambda-\lambda_{1}), \notag\\
z=T_{2}(\lambda)=\sqrt{4t(1-48\delta\lambda_{2}^{2})}(\lambda-\lambda_{2}), \notag\\
z=T_{3}(\lambda)=\sqrt{4t(48\delta\lambda_{3}^{2}-1)}(\lambda-\lambda_{3}).
\end{align}
For instance, after the scaling transformation,  a function $\varphi(\lambda)$ becomes
\begin{align}
T_{1}(\varphi(\lambda))=\varphi(\frac{z}{\sqrt{4t(48\delta\lambda_{1}^{2}-1)}}+\lambda_{1}), \notag\\ T_{2}(\varphi(\lambda))=\varphi(\frac{z}{\sqrt{4t(1-48\delta\lambda_{2}^{2})}}+\lambda_{2}), \notag\\ T_{3}(\varphi(\lambda))=\varphi(\frac{z}{\sqrt{4t(48\delta\lambda_{3}^{2}-1)}}+\lambda_{3}).
\end{align}
Hence, one has
\begin{align}\label{32}
T_{1}(e^{itf}\delta(\lambda))
=\delta_{\lambda_{1}}^{0}\delta_{\lambda_{1}}^{1}(z),\quad
T_{2}(e^{itf}\delta(\lambda))
=\delta_{\lambda_{2}}^{0}\delta_{\lambda_{2}}^{1}(z),\quad
T_{3}(e^{itf}\delta(\lambda))
=\delta_{\lambda_{3}}^{0}\delta_{\lambda_{3}}^{1}(z),
\end{align}
where
\begin{align}\label{33}
\delta_{\lambda_{1}}^{0}&=\left[4t(48\delta\lambda_{1}^{2}-1)\right]^{\frac{i\vartheta(\lambda_{1})}{2}}(\lambda_{1}-\lambda_{3})^{-i\vartheta(\lambda_{3})}
(\lambda_{2}-\lambda_{1})^{i\vartheta(\lambda_{1})}
e^{i\lambda_{1}^{2}t(24\delta\lambda_{1}^{2}-1)+\chi_{1}(\lambda_{1})},\notag\\
\delta_{\lambda_{1}}^{1}&=(-z)^{-i\vartheta(\lambda_{1})}\left(\frac{z/\sqrt{4t(48\delta\lambda_{1}^{2}-1)}+\lambda_{1}-\lambda_{3}}{\lambda_{1}-
\lambda_{3}}\right)
^{-i\vartheta(\lambda_{3})}\left(\frac{\lambda_{1}-
\lambda_{2}}{z/\sqrt{4t(48\delta\lambda_{1}^{2}-1)}+\lambda_{1}-\lambda_{2}}\right)
^{-i\vartheta(\lambda_{1})}\notag\\
&e^{-\frac{i\delta z^{4}}{2t(48\delta\lambda_{1}^{2}-1)^{2}}-\frac{32it\delta\lambda_{1}z^{3}}{(4t(48\delta\lambda_{1}^{2}-1))^{3/2}}-\frac{i}{4}z^{2}}
e^{\chi_{1}(\frac{z}{\sqrt{4t(48\delta\lambda_{1}^{2}-1)}}+\lambda_{1})-\chi_{1}(\lambda_{1})},\notag\\
\delta_{\lambda_{2}}^{0}&=\left[4t(1-48\delta\lambda_{2}^{2})\right]^{-\frac{i\vartheta(\lambda_{2})}{2}}(\lambda_{2}-\lambda_{1})^{-i\vartheta(\lambda_{1})}
(\lambda_{2}-\lambda_{3})^{-i\vartheta(\lambda_{2})}
e^{i\lambda_{2}^{2}t(24\delta\lambda_{2}^{2}-1)+\chi_{2}(\lambda_{2})},\notag\\
\delta_{\lambda_{2}}^{1}&=z^{i\vartheta(\lambda_{2})}\left(\frac{-z/\sqrt{4t(1-48\delta\lambda_{2}^{2})}-\lambda_{2}+\lambda_{1}}{\lambda_{1}-
\lambda_{2}}\right)
^{-i\vartheta(\lambda_{1})}\left(\frac{z/\sqrt{4t(1-48\delta\lambda_{2}^{2})}+\lambda_{2}-\lambda_{3}}{\lambda_{2}-
\lambda_{3}}\right)
^{-i\vartheta(\lambda_{1})}\notag\\
&e^{-\frac{i\delta z^{4}}{2t(1-48\delta\lambda_{2}^{2})^{2}}-\frac{32it\delta\lambda_{2}z^{3}}{(4t(1-48\delta\lambda_{2}^{2}))^{3/2}}+\frac{i}{4}z^{2}}
e^{\chi_{2}(\frac{z}{\sqrt{4t(1-48\delta\lambda_{2}^{2})}}+\lambda_{2})-\chi_{2}(\lambda_{2})},\notag\\
\delta_{\lambda_{3}}^{0}&=\left[4t(48\delta\lambda_{3}^{2}-1)\right]^{\frac{i\vartheta(\lambda_{3})}{2}}(\lambda_{3}-\lambda_{1})^{-i\vartheta(\lambda_{1})}
(\lambda_{2}-\lambda_{3})^{i\vartheta(\lambda_{3})}
e^{i\lambda_{3}^{2}t(24\delta\lambda_{3}^{2}-1)+\chi_{3}(\lambda_{3})},\notag\\
\delta_{\lambda_{3}}^{1}&=(-z)^{-i\vartheta(\lambda_{3})}\left(\frac{z/\sqrt{4t(48\delta\lambda_{3}^{2}-1)}+\lambda_{3}-\lambda_{1}}{\lambda_{3}-
\lambda_{1}}\right)
^{-i\vartheta(\lambda_{1})}\left(\frac{\lambda_{2}-
\lambda_{3}}{-z/\sqrt{4t(48\delta\lambda_{3}^{2}-1)}+\lambda_{2}-\lambda_{3}}\right)
^{-i\vartheta(\lambda_{3})}\notag\\
&e^{-\frac{i\delta z^{4}}{2t(48\delta\lambda_{3}^{2}-1)^{2}}-\frac{32it\delta\lambda_{3}z^{3}}{(4t(48\delta\lambda_{3}^{2}-1))^{3/2}}-\frac{i}{4}z^{2}}
e^{\chi_{3}(\frac{z}{\sqrt{4t(48\delta\lambda_{3}^{2}-1)}}+\lambda_{3})-\chi_{3}(\lambda_{3})}.
\end{align}

\begin{figure}[!htb]
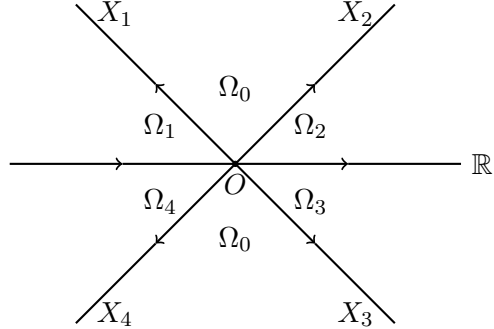

\centering
\caption{\footnotesize The jump contour $X$  and domains $\Omega_{j}(j=0,\cdots, 4)$.}
\label{F4}
\end{figure}

Let $t\rightarrow\infty$, the functions appeared in the \eqref{33} have the following approaches
\begin{align}\label{34}
&r_{j}(\lambda(z))\rightarrow r_{j}(\lambda_{l}),\quad j=1,2, \quad l=1,2, 3,\notag\\
&\delta_{\lambda_{l}}^{1}\rightarrow e^{(-1)^{l}\frac{i}{4}z^{2}}((-1)^{l}z)^{(-1)^{l}i\vartheta(\lambda_{l})}.
\end{align}
This implies that we can derive the following RH problem $M^{X_{l}}(l=1, 2, 3)$ as $t\rightarrow\infty$ in the $z$ plane relatived to $X=X_{1}\cup X_{2}\cup X_{3}\cup X_{4}$ (see Figure \ref{F4})
\begin{align}\label{35}
\left\{
\begin{array}{lr}
M^{X_{l}}_{+}(\xi, z)=M^{X_{l}}_{-}(\xi, z)J^{X_{l}}(\xi, z), \qquad z\in X,\\
M^{X_{l}}(\xi, z)\rightarrow I,\qquad z\rightarrow \infty,
  \end{array}
\right.
\end{align}
where  the jump matrix is
\begin{align}\label{36}
J^{X_{1}}=\left\{
\begin{array}{lr}
\left(\begin{array}{cc}
    1 &  0\\
    r_{1}(\lambda_{1})e^{\frac{i}{2}z^{2}}(-z)^{2i\vartheta(\lambda_{1})}  &  1\\
\end{array}\right), \qquad \qquad z\in X_{1},\\
\left(\begin{array}{cc}
    1 &  \frac{r_{2}(\lambda_{1})}{1-r_{1}(\lambda_{1})r_{2}(\lambda_{1})}e^{-\frac{i}{2}z^{2}}(-z)^{-2i\vartheta(\lambda_{1})}\\
    0  &  1\\
\end{array}\right),  \qquad\quad\ z\in X_{2},\\
\left(\begin{array}{cc}
    1 &  0\\
    -\frac{r_{1}(\lambda_{1})}{1-r_{1}(\lambda_{1})r_{2}(\lambda_{1})}e^{\frac{i}{2}z^{2}}(-z)^{2i\vartheta(\lambda_{1})}  &  1\\
\end{array}\right), \qquad z\in X_{3},\\
\left(\begin{array}{cc}
    1 &  -r_{2}(\lambda_{1})e^{-\frac{i}{2}z^{2}}(-z)^{-2i\vartheta(\lambda_{1})}\\
    0  &  1\\
\end{array}\right), \qquad z\in X_{4}.
  \end{array}
\right.\notag\\
J^{X_{2}}=\left\{
\begin{array}{lr}
\left(\begin{array}{cc}
    1 &  -\frac{r_{2}(\lambda_{2})}{1-r_{1}(\lambda_{2})r_{2}(\lambda_{2})}e^{\frac{i}{2}z^{2}}z^{2i\vartheta(\lambda_{2})}\\
    0  &  1\\
\end{array}\right), \qquad \qquad z\in X_{1},\\
\left(\begin{array}{cc}
    1 &  0\\
    -r_{1}(\lambda_{2})e^{-\frac{i}{2}z^{2}}z^{-2i\vartheta(\lambda_{2})}  &  1\\
\end{array}\right),  \qquad\quad\ z\in X_{2},\\
\left(\begin{array}{cc}
    1 &  r_{2}(\lambda_{2})e^{\frac{i}{2}z^{2}}z^{2i\vartheta(\lambda_{2})}\\
    0  &  1\\
\end{array}\right), \qquad z\in X_{3},\\
\left(\begin{array}{cc}
    1 &  0\\
    \frac{r_{1}(\lambda_{2})}{1-r_{1}(\lambda_{2})r_{2}(\lambda_{2})}e^{-\frac{i}{2}z^{2}}z^{-2i\vartheta(\lambda_{2})}  &  1\\
\end{array}\right), \qquad z\in X_{4}.
  \end{array}
\right.\notag\\
J^{X_{3}}=\left\{
\begin{array}{lr}
\left(\begin{array}{cc}
    1 &  0\\
    r_{1}(\lambda_{3})e^{\frac{i}{2}z^{2}}(-z)^{2i\vartheta(\lambda_{3})}  &  1\\
\end{array}\right), \qquad \qquad z\in X_{1},\\
\left(\begin{array}{cc}
    1 &  \frac{r_{2}(\lambda_{3})}{1-r_{1}(\lambda_{3})r_{2}(\lambda_{3})}e^{-\frac{i}{2}z^{2}}(-z)^{-2i\vartheta(\lambda_{3})}\\
    0  &  1\\
\end{array}\right),  \qquad\quad\ z\in X_{2},\\
\left(\begin{array}{cc}
    1 &  0\\
    -\frac{r_{1}(\lambda_{3})}{1-r_{1}(\lambda_{3})r_{2}(\lambda_{3})}e^{\frac{i}{2}z^{2}}(-z)^{2i\vartheta(\lambda_{3})}  &  1\\
\end{array}\right), \qquad z\in X_{3},\\
\left(\begin{array}{cc}
    1 &  -r_{2}(\lambda_{3})e^{-\frac{i}{2}z^{2}}(-z)^{-2i\vartheta(\lambda_{3})}\\
    0  &  1\\
\end{array}\right), \qquad z\in X_{4}.
  \end{array}
\right.
\end{align}
\\
\centerline{\begin{tikzpicture}
\draw[fill] (0,0) circle [radius=0.035];
\draw[-][thick](-2.12,-2.12)--(-1.06,-1.06);
\draw[<-][thick](-1.06,-1.06)--(0,0)node[below]{$O$};
\draw[->][thick](0,0)--(1.06,1.06);
\draw[-][thick](1.06,1.06)--(2.12,2.12);
\draw[-][thick](-2.12,2.12)--(-1.06,1.06);
\draw[<-][thick](-1.06,1.06)--(0,0);
\draw[->][thick](0,0)--(1.06,-1.06);
\draw[->][thick](-3,0)--(-1.5,0);
\draw[->][thick](-1.5,0)--(1.5,0);
\draw[-][thick](1.5,0)--(3.0,0)node[right]{$\mathbb{R}$};
\draw[-][thick](1.06,-1.06)--(2.12,-2.12);
\draw[fill] (0,1) node{$\Omega_{0}$};
\draw[fill] (0,-1) node{$\Omega_{0}$};
\draw[fill] (-1,0.5) node{$\Omega_{1}$};
\draw[fill] (1,0.5) node{$\Omega_{2}$};
\draw[fill] (-1,-0.5) node{$\Omega_{4}$};
\draw[fill] (1,-0.5) node{$\Omega_{3}$};
\draw[fill] (-1.6,2.0) node{$X_{1}$};
\draw[fill] (-1.6,-2.0) node{$X_{4}$};
\draw[fill] (1.6,2.0) node{$X_{2}$};
\draw[fill] (1.6,-2.0) node{$X_{3}$};
\end{tikzpicture}}\\

For $l=1, 2, 3$, defining $D_{\epsilon}(\lambda_{l})$ as the open disk of radius $\epsilon$ centered at $\lambda_{l}$ for a small
$\epsilon>0$ and using $M^{X_{l}}$, we introduce $M^{\lambda_{l}}$ for $\lambda \in D_{\epsilon}(\lambda_{l})$
\begin{align}\label{37}
M^{\lambda_{l}}(x, t, \lambda)=(\delta_{\lambda_{l}}^{0})^{\sigma_{3}}M^{X_{l}}(z)(\delta_{\lambda_{l}}^{0})^{-\sigma_{3}},
\end{align}
which is analytic function in region $\lambda\in D_{\epsilon}(\lambda_{l})\setminus X_{\lambda_{l}}^{\epsilon}$, where $X_{\lambda_{l}}^{\epsilon}=X_{\lambda_{l}}\cap D_{\epsilon}(\lambda_{l}), X_{\lambda_{l}}=X+\lambda_{l}$ means the cross $X$  centered at $\lambda_{l}$. It is not hard to find that $M^{\lambda_{l}}(x,t,\lambda)$ solves following  RH problem, given by $M_{+}^{\lambda_{l}}(x,t,\lambda)=M_{-}^{\lambda_{l}}(x,t,\lambda)J^{\lambda_{l}}$.

Let $A^{(j)}$  represents the $j$th column of a matrix $A$, we obtain the following estimates for
the jump matrix $J^{\lambda_{l}}$ as $1\leq n\leq\infty$:
\begin{align}\label{38}
\|(\tilde{\tilde{J}}-J^{\lambda_{l}})^{(j)}\|_{L^{n}(X_{\lambda_{l}}^{\epsilon})}\leq Ct^{-\frac{1}{2}-\frac{1}{2n}+(-1)^{l+j}\mbox{Im}\vartheta(\lambda_{l})}\ln t,\quad j=1, 2,
\end{align}
where $C$ is a positive constant independent of $t,\xi,\lambda$. Furthermore, as $t\rightarrow\infty$,
\begin{align}\label{39}
\|((M^{\lambda_{l}})^{-1}-I)^{(j)}\|_{L^{\infty}(\partial D_{\epsilon}(\lambda_{l}))}\leq Ct^{-\frac{1}{2}+(-1)^{l+j}\mbox{Im}\vartheta(\lambda_{l})},
\end{align}
and
\begin{align}\label{40}
\frac{1}{2\pi i}\int_{\partial D_{\epsilon}(\lambda_{l})}(M^{\lambda_{l}})^{-1}-I)^{(j)}d\lambda=-\frac{((\delta_{\lambda_{l}}^{0})
^{\hat{\sigma}_{3}}M_{1}^{X_{l}})^{(j)}}{\sqrt{4t(-1)^{l}(1-48\delta\lambda_{l}^{2})}}+O(t^{-1+(-1)^{l+j}\mbox{Im}\vartheta(\lambda_{l})}),
\end{align}
where $M_{1}^{X_{l}}$ is given by the asymptotic expansion
\begin{align}\label{41}
M^{X_{l}}(x, t, z)=I+\frac{M_{1}^{X_{l}}(x, t)}{z}+O(\frac{1}{z^{2}}),\qquad z\rightarrow \infty.
\end{align}

Next, we devote to derive the explicit expression of long-time asymptotic behavior for the nonlocal LPD equation \eqref{1} on the line by introducing the approximate solution $M^{(app)}(x,t,\lambda)$ as follows
\begin{align}\label{42}
M^{app}=\left\{
\begin{array}{lr}
M^{\lambda_{1}}, \qquad  \lambda\in D_{\epsilon}(\lambda_{1}),\\
M^{\lambda_{2}}, \qquad  \lambda\in D_{\epsilon}(\lambda_{2}),\\
M^{\lambda_{3}}, \qquad  \lambda\in D_{\epsilon}(\lambda_{3}),\\
I,  \qquad\quad\ \mbox{elsewhere}.\\
  \end{array}
\right.
\end{align}
Defining $\hat{M}(x,t,\lambda)=\tilde{\tilde{M}}(M^{app})^{-1}$, then we get the following RH problem for $\hat{M}(x,t,\lambda)$
\begin{align}\label{43}
\hat{M}_{+}(x,t,\lambda)=\hat{M}_{-}(x,t,\lambda)\hat{J}(x,t,\lambda),\quad \lambda\in \hat{\Sigma},
\end{align}
where the jump contour $\hat{\Sigma}=\Sigma\cup \partial D_{\epsilon}(\lambda_{1})\cup \partial D_{\epsilon}(\lambda_{2})\cup \partial D_{\epsilon}(\lambda_{3})$, and the jump matrix $\hat{J}(x,t, \lambda)$ arrives at
\begin{align}\label{44}
\hat{J}=\left\{
\begin{array}{lr}
M_{-}^{app}\tilde{\tilde{J}}(M_{+}^{app})^{-1}, \qquad  \lambda\in \hat{\Sigma}\cap ( D_{\epsilon}(\lambda_{1})\cup D_{\epsilon}(\lambda_{2})\cup D_{\epsilon}(\lambda_{3})),\\
(M^{app})^{-1}, \qquad  \lambda\in \partial D_{\epsilon}(\lambda_{1})\cup \partial D_{\epsilon}(\lambda_{2})\cup \partial D_{\epsilon}(\lambda_{3}),\\
\tilde{\tilde{J}},  \qquad\quad\ \lambda\in \hat{\Sigma}\setminus (\overline{ D_{\epsilon}(\lambda_{1})}\cup \overline{ D_{\epsilon}(\lambda_{2})}\cup \overline{ D_{\epsilon}(\lambda_{3})}).\\
  \end{array}
\right.
\end{align}

Let $\hat{\omega}=\hat{J}-I$, for $1\leq n\leq\infty$, $\hat{\Sigma}_{1}=\bigcup_{1}^{4}\gamma_{k}\setminus (D_{\epsilon}(\lambda_{1})\cup D_{\epsilon}(\lambda_{2})\cup D_{\epsilon}(\lambda_{3}))$ and $\xi\in (-\sqrt{\frac{1}{27\delta}}+\epsilon, \sqrt{\frac{1}{27\delta}}-\epsilon)$, according to \eqref{38}, \eqref{39}, \eqref{42} and \eqref{44}, we have the following estimates, given by
\begin{gather}
\|\hat{\omega}\|_{L^{n}(\hat{\Sigma}_{1})}\leq Ce^{-ct},\notag\\
\|\hat{\omega}^{(j)}\|_{L^{n}(\partial D_{\epsilon}(\lambda_{l}))}\leq Ct^{-\frac{1}{2}+(-1)^{l+j}\mbox{Im}\vartheta(\lambda_{l})},\notag\\
\|\hat{\omega}^{(j)}\|_{L^{n}(X_{\lambda_{l}}^{\epsilon})}\leq Ct^{-\frac{1}{2}-\frac{1}{2n}+(-1)^{l+j}\mbox{Im}\vartheta(\lambda_{l})}\ln t.\label{45}
\end{gather}

Here, we shall define the Cauchy operator
$(\mathcal{C}f)(\lambda)=\frac{1}{2\pi i}\int_{\hat{\Sigma}}\frac{f(s)}{s-\lambda}\mathrm{d}s, \lambda\in \mathbb{C}\setminus\hat{\Sigma},$
and the integral operator $\mathcal{C}_{\hat{\omega}}:L^{2}(\hat{\Sigma})+L^{\infty}(\hat{\Sigma})\rightarrow L^{2}(\hat{\Sigma})$ by $\mathcal{C}_{\hat{\omega}}f=\mathcal{C}_{-}(f\hat{\omega})$, then, we have
\begin{align}\label{47}
\|\mathcal{C}_{\hat{\omega}}\|_{B(L^{2}(\hat{\Sigma}))}\leq C\|\hat{\omega}\|_{L^{\infty}(\hat{\Sigma})}\leq C t^{-\frac{1}{2}+max\{|\mbox{Im}\vartheta(\lambda_{1})|,|\mbox{Im}\vartheta(\lambda_{2})|,|\mbox{Im}\vartheta(\lambda_{3})|\}}\ln t,\ t\rightarrow\infty,
\end{align}
where $B(L^{2}(\hat{\Sigma}))$ means the Banach space of bounded linear operators $L^{2}(\hat{\Sigma})\rightarrow L^{2}(\hat{\Sigma})$.
Moreover,
\begin{align}\label{48}
\|\hat{\mu}-I\|_{L^{2}(\hat{\Sigma})}\leq C\|\hat{\omega}\|_{L^{2}(\hat{\Sigma})},
\end{align}
where $\hat{\mu}=I+L^{2}(\hat{\Sigma})$ satisfies the following integral equation
\begin{align}\label{49}
\hat{\mu}=I+\mathcal{C}_{\hat{\omega}}\hat{\mu}.
\end{align}
Finally, the representation for $\hat{M}$ admits
\begin{align}\label{50}
\hat{M}(x,t,\lambda)=I+\frac{1}{2\pi i}\int_{\hat{\Sigma}}\frac{\hat{\mu}(x,t,s)\hat{\omega}(x,t,s)}{s-\lambda}\mathrm{d}s,
\end{align}
it follows that
\begin{align}\label{51}
\lim_{\lambda\rightarrow\infty}\lambda(\hat{M}(x,t,\lambda)-I)=-\frac{1}{2\pi i}\int_{\hat{\Sigma}}\hat{\mu}(x,t,\lambda)\hat{\omega}(x,t,\lambda)\mathrm{d}\lambda.
\end{align}
Using \eqref{45} and \eqref{48}, one has
\begin{align}\label{52}
\int_{\hat{\Sigma}_{1}}\hat{\mu}(x,t,\lambda)\hat{\omega}(x,t,\lambda)\mathrm{d}\lambda&=
\int_{\hat{\Sigma}_{1}}\hat{\omega}(x,t,\lambda)\mathrm{d}\lambda
+\int_{\hat{\Sigma}_{1}}(\hat{\mu}(x,t,\lambda)-I)\hat{\omega}(x,t,\lambda)\mathrm{d}\lambda\notag\\
&\leq \|\hat{\omega}\|_{L^{1}(\hat{\Sigma}_{1})}+
\|\hat{\mu}-I\|_{L^{2}(\hat{\Sigma}_{1})}\|\hat{\omega}\|_{L^{2}(\hat{\Sigma}_{1})}\notag\\
&\leq Ce^{-ct}\leq Ct^{-1},\qquad t\rightarrow \infty.
\end{align}
\begin{align}\label{53}
\int_{X_{\lambda_{l}}^{\epsilon}}(\hat{\mu}(x,t,\lambda)\hat{\omega}(x,t,\lambda))^{(j)}\mathrm{d}\lambda&=
\int_{X_{\lambda_{l}}^{\epsilon}}(\hat{\omega}(x,t,\lambda))^{(j)}\mathrm{d}\lambda
+\int_{X_{\lambda_{l}}^{\epsilon}}((\hat{\mu}(x,t,\lambda)-I)\hat{\omega}(x,t,\lambda))^{(j)}\mathrm{d}\lambda\notag\\
&\leq \|\hat{\omega}^{(j)}\|_{L^{1}(X_{\lambda_{l}}^{\epsilon})}+
\|(\hat{\mu}-I)^{(j)}\|_{L^{2}(X_{\lambda_{l}}^{\epsilon})}\|\hat{\omega}^{(j)}\|_{L^{2}(X_{\lambda_{l}}^{\epsilon})}\notag\\
&\leq Ct^{-1+(-1)^{l+j}\mbox{Im}\vartheta(\lambda_{l})}\ln t,\qquad t\rightarrow \infty.
\end{align}
\begin{align}\label{54}
&-\frac{1}{2\pi i}\int_{\partial D_{\epsilon}(\lambda_{l})}(\hat{\mu}(x,t,\lambda)\hat{\omega}(x,t,\lambda))^{(j)}\mathrm{d}\lambda\notag\\
&=
-\frac{1}{2\pi i}\int_{\partial D_{\epsilon}(\lambda_{l})}\hat{\omega}(x,t,\lambda)^{(j)}\mathrm{d}\lambda
-\frac{1}{2\pi i}\int_{\partial D_{\epsilon}(\lambda_{l})}((\hat{\mu}(x,t,\lambda)-I)\hat{\omega}(x,t,\lambda))^{(j)}\mathrm{d}\lambda\notag\\
&\leq -\frac{1}{2\pi i}\int_{\partial D_{\epsilon}(\lambda_{l})}((M^{\lambda_{l}})^{-1}-I)^{(j)}\mathrm{d}\lambda+
O(\|(\hat{\mu}-I)^{(j)}\|_{L^{2}(\partial D_{\epsilon}(\lambda_{l}))}\|\hat{\omega}^{(j)}\|_{L^{2}(\partial D_{\epsilon}(\lambda_{l}))})\notag\\
&\leq \frac{((\delta_{\lambda_{l}}^{0})
^{\hat{\sigma}_{3}}M_{1}^{X_{l}})^{(j)}}{\sqrt{4t(-1)^{l}(1-48\delta\lambda_{l}^{2})}}+Q(\xi,t),\quad t\rightarrow \infty,
\end{align}
where
\begin{align}\label{55}
Q(\xi,t)=\left\{
\begin{array}{lr}
O(t^{-1+2|\mbox{Im}\vartheta(\lambda_{l})|}),  \qquad (-1)^{l+j}\mbox{Im}\vartheta(\lambda_{l})\geq 0,\\
O(t^{-1-|\mbox{Im}\vartheta(\lambda_{l})|}),  \qquad (-1)^{l+j}\mbox{Im}\vartheta(\lambda_{l})<0.\\
  \end{array}
\right.
\end{align}
Therefore, we derive the following important result
\begin{align}\label{56}
&q(x,t)=-2i\lim_{\lambda\rightarrow\infty}\lambda[M(x,t,\lambda)]_{12}=-2i\lim_{\lambda\rightarrow\infty}\lambda[\hat{M}(x,t,\lambda)-I]_{12}\notag\\
&=\frac{1}{\pi}\int_{\hat{\Sigma}}[\hat{\mu}(x,t,\lambda)\hat{\omega}(x,t,\lambda)]_{12}\mathrm{d}\lambda
=\frac{1}{\pi}\left(\int_{\hat{\Sigma}_{1}}[\hat{\mu}(x,t,\lambda)\hat{\omega}(x,t,\lambda)]_{12}\mathrm{d}\lambda\right.\notag\\
&\left.+\sum_{l=1}^{3}\int_{X_{\lambda_{l}}^{\epsilon}}[\hat{\mu}(x,t,\lambda)\hat{\omega}(x,t,\lambda)]_{12}\mathrm{d}\lambda
+\sum_{l=1}^{3}\int_{\partial D_{\epsilon}(\lambda_{l})}[\hat{\mu}(x,t,\lambda)\hat{\omega}(x,t,\lambda)]_{12}\mathrm{d}\lambda\right)\notag\\
&=-2i\sum_{l=1}^{3}\frac{(\delta_{\lambda_{l}}^{0})^{2}[M_{1}^{X_{l}}]_{12}}{\sqrt{4t(-1)^{l}(1-48\delta\lambda_{l}^{2})}}+R(\xi,t),\qquad t\rightarrow \infty,
\end{align}
where
\begin{align}\label{57}
R(\xi,t)=\left\{
\begin{array}{lr}
O(t^{-1+2max\{|\mbox{Im}\vartheta(\lambda_{1})|,|\mbox{Im}\vartheta(\lambda_{2})|,|\mbox{Im}\vartheta(\lambda_{3})|\}}),  \quad (-1)^{l}\mbox{Im}\vartheta(\lambda_{l})> 0,\\
O(t^{-1+2max\{|\mbox{Im}\vartheta(\lambda_{1})|,|\mbox{Im}\vartheta(\lambda_{2})|\}}),  \quad \mbox{Im}\vartheta(\lambda_{1})<0,\mbox{Im}\vartheta(\lambda_{2})>0,\mbox{Im}\vartheta(\lambda_{3})\geq 0,\\
O(t^{-1+2max\{|\mbox{Im}\vartheta(\lambda_{2})|,|\mbox{Im}\vartheta(\lambda_{3})|\}}),  \quad \mbox{Im}\vartheta(\lambda_{1})\geq 0,\mbox{Im}\vartheta(\lambda_{2})>0,\mbox{Im}\vartheta(\lambda_{3})<0,\\
O(t^{-1+2max\{|\mbox{Im}\vartheta(\lambda_{1})|,|\mbox{Im}\vartheta(\lambda_{3})|\}}),  \quad \mbox{Im}\vartheta(\lambda_{1})<0,\mbox{Im}\vartheta(\lambda_{2})\leq 0,\mbox{Im}\vartheta(\lambda_{3})< 0,\\
O(t^{-1+2|\mbox{Im}\vartheta(\lambda_{1})|}),  \quad\qquad\qquad\qquad \mbox{Im}\vartheta(\lambda_{1})<0,\mbox{Im}\vartheta(\lambda_{2})\leq 0,\mbox{Im}\vartheta(\lambda_{3})\geq 0,\\
O(t^{-1+2|\mbox{Im}\vartheta(\lambda_{2})|}),  \quad\qquad\qquad\qquad \mbox{Im}\vartheta(\lambda_{1})\geq 0,\mbox{Im}\vartheta(\lambda_{2})> 0,\mbox{Im}\vartheta(\lambda_{3})\geq 0,\\
O(t^{-1+2|\mbox{Im}\vartheta(\lambda_{3})|}),  \quad\qquad\qquad\qquad \mbox{Im}\vartheta(\lambda_{1})\geq 0,\mbox{Im}\vartheta(\lambda_{2})\leq 0,\mbox{Im}\vartheta(\lambda_{3})< 0,\\
O(t^{-1}\ln t),  \qquad\qquad\quad \mbox{Im}\vartheta(\lambda_{l})=0, (-1)^{s}\mbox{Im}\vartheta(\lambda_{s})\leq 0, s=1,2,3\ \mbox{and}\ s \neq l, \\
O(t^{-1}),  \qquad\qquad\qquad\qquad\qquad\qquad\qquad\qquad (-1)^{l}\mbox{Im}\vartheta(\lambda_{l})< 0.\\
  \end{array}
\right.
\end{align}

\subsection{Solving the model RH problem}
In this subsection, we focus on solving the model RH problem by giving $[M_{1}^{X_{1}}]_{12}$ explicitly. To realize it, we
introduce the following transformation(see Figure \ref{F4})
\begin{align}\label{58}
M^{mod}=M^{X_{1}}G_{j},\qquad  z \in \Omega_{j},\qquad j=0, \cdots, 4,
\end{align}
where
\begin{gather}\label{59}
G_{0}=e^{-\frac{1}{4}iz^{2}\sigma_{3}}(-z)^{-i\vartheta(\lambda_{1})\sigma_{3}},\notag\\
G_{1}=G_{0}\left(\begin{array}{cc}
    1  &  0\\
    -r_{1}(\lambda_{1})  &  1\\
\end{array}\right),\quad G_{2}=G_{0}\left(\begin{array}{cc}
    1  &  \frac{r_{2}(\lambda_{1})}{1-r_{1}(\lambda_{1})r_{2}(\lambda_{1})}\\
    0  &  1\\
\end{array}\right),\notag\\
G_{3}=G_{0}\left(\begin{array}{cc}
    1  &  0\\
    \frac{r_{1}(\lambda_{1})}{1-r_{1}(\lambda_{1})r_{2}(\lambda_{1})}  &  1\\
\end{array}\right),\quad G_{4}=G_{0}\left(\begin{array}{cc}
    1  &  -r_{2}(\lambda_{1})\\
    0  &  1\\
\end{array}\right).
\end{gather}
In terms of this transformation, a model RH problem for $M^{mod}$ with a constant jump matrix is derived
\begin{align}\label{60}
\left\{
\begin{array}{lr}
M^{mod}(z)\ \mbox{is analytic in} \ \mathbb{C}\backslash \mathbb{R},\\
M^{mod}_{+}(z)=M^{mod}_{-}(z)J^{mod}(\lambda_{1}), \quad z\in \mathbb{R},\\
M^{mod}(z)\rightarrow e^{-\frac{1}{4}iz^{2}\sigma_{3}}(-z)^{-i\vartheta(\lambda_{1})\sigma_{3}},\qquad z\rightarrow \infty,
  \end{array}
\right.
\end{align}
where
\begin{align}
J^{mod}(\lambda_{1})=\left(\begin{array}{cc}
    1-r_{1}(\lambda_{1})r_{2}(\lambda_{1})  &  r_{2}(\lambda_{1})\\
  -r_{1}(\lambda_{1}) &  1\\
\end{array}\right).
\end{align}
Using the Liouville's theorem and parabolic cylinder functions, we can explicitly solve this RH problem. The jump matrix $J^{mod}$  is constant, which indicates that $\frac{\mathrm{d}}{\mathrm{d}z}M^{mod}(M^{mod})^{-1}$ possesses continuous jump along any of the rays. Then, one has
\begin{align}\label{61}
\frac{\mathrm{d}}{\mathrm{d}z}M^{mod}+\left(\begin{array}{cc}
    \frac{i}{2}z  &  \beta\\
  \alpha &  -\frac{i}{2}z\\
\end{array}\right)M^{mod}=0,
\end{align}
where $\beta=-i\left[M_{1}^{X_{1}}\right]_{12}, \alpha=i\left[M_{1}^{X_{1}}\right]_{21}$.
The solution of $\eqref{61}$ arrives at
\begin{align}\label{62}
M^{mod}=\left(\begin{array}{cc}
    M_{11}^{mod}  &  \frac{\frac{i}{2}zM_{22}^{mod}-\frac{\mathrm{d}M_{22}^{mod}}{\mathrm{d}z}}{\alpha}\\
  \frac{\frac{i}{2}zM_{11}^{mod}+\frac{\mathrm{d}M_{11}^{mod}}{\mathrm{d}z}}{-\beta} &  M_{22}^{mod}\\
\end{array}\right),
\end{align}
where the functions $M_{jj}^{mod}, j=1, 2,$ solve the following equations
\begin{align}\label{63}
&\frac{\mathrm{d}^{2}}{\mathrm{d}z^{2}}M_{11}^{mod}+(\frac{i}{2}-\alpha\beta+\frac{z^{2}}{4})M_{11}^{mod}=0,\notag\\
&\frac{\mathrm{d}^{2}}{\mathrm{d}z^{2}}M_{22}^{mod}+(-\frac{i}{2}-\alpha\beta+\frac{z^{2}}{4})M_{22}^{mod}=0.
\end{align}
Since the above equations are standard parabolic cylinder equation and $M_{11}^{mod}\rightarrow e^{-\frac{1}{4}iz^{2}}(-z)^{-i\vartheta(\lambda_{1})},$ $ M_{22}^{mod}\rightarrow e^{\frac{1}{4}iz^{2}}(-z)^{i\vartheta(\lambda_{1})}, z\rightarrow\infty$, we get
\begin{align}\label{64}
M_{11}^{mod}=\left\{
\begin{array}{lr}
(-e^{-\frac{3\pi}{4}i})^{i\vartheta(\lambda_{1})}D_{-i\vartheta(\lambda_{1})}(e^{-\frac{3\pi}{4}i}z) \qquad \mbox{Im}(z)>0,\\
\\
(-e^{\frac{\pi}{4}i})^{i\vartheta(\lambda_{1})}D_{-i\vartheta(\lambda_{1})}(e^{\frac{\pi}{4}i}z) \qquad \qquad \mbox{Im}(z)<0,
  \end{array}
\right.
\end{align}
\begin{align}\label{65}
M_{22}^{mod}=\left\{
\begin{array}{lr}
(-e^{-\frac{\pi}{4}i})^{-i\vartheta(\lambda_{1})}D_{i\vartheta(\lambda_{1})}(e^{-\frac{\pi}{4}i}z) \ \qquad \mbox{Im}(z)>0,\\
\\
(-e^{\frac{3\pi}{4}i})^{-i\vartheta(\lambda_{1})}D_{i\vartheta(\lambda_{1})}(e^{\frac{3\pi}{4}i}z) \quad \qquad \mbox{Im}(z)<0.
  \end{array}
\right.
\end{align}
Then, one has
\begin{gather}
M^{mod}_{-}(z)^{-1}M^{mod}_{+}(z)=M^{mod}_{-}(0)^{-1}M^{mod}_{+}(0)=\notag\\
\left(\begin{array}{cc}
  (-e^{\frac{\pi}{4}i})^{i\vartheta}\frac{2^{\frac{-i\vartheta}{2}}\sqrt{\pi}}{\Gamma(\frac{1+i\vartheta}{2})}    & e^{\frac{3\pi}{4}i}(-e^{\frac{3\pi}{4}i})^{-i\vartheta}\frac{2^{\frac{1+i\vartheta}{2}}\sqrt{\pi}}{\alpha\Gamma(\frac{-i\vartheta}{2})} \\
  e^{\frac{\pi}{4}i}(-e^{\frac{\pi}{4}i})^{i\vartheta}\frac{2^{\frac{1-i\vartheta}{2}}\sqrt{\pi}}{\beta\Gamma(\frac{i\vartheta}{2})} &   (-e^{\frac{3\pi}{4}i})^{-i\vartheta}\frac{2^{\frac{i\vartheta}{2}}\sqrt{\pi}}{\Gamma(\frac{1-i\vartheta}{2})}\\
\end{array}\right)^{-1}\notag\\
\left(\begin{array}{cc}
  (-e^{-\frac{3\pi}{4}i})^{i\vartheta}\frac{2^{\frac{-i\vartheta}{2}}\sqrt{\pi}}{\Gamma(\frac{1+i\vartheta}{2})}    & e^{-\frac{\pi}{4}i}(-e^{-\frac{\pi}{4}i})^{-i\vartheta}\frac{2^{\frac{1+i\vartheta}{2}}\sqrt{\pi}}{\alpha\Gamma(\frac{-i\vartheta}{2})} \\
  e^{-\frac{3\pi}{4}i}(-e^{-\frac{3\pi}{4}i})^{i\vartheta}\frac{2^{\frac{1-i\vartheta}{2}}\sqrt{\pi}}{\beta\Gamma(\frac{i\vartheta}{2})} &   (-e^{-\frac{\pi}{4}i})^{-i\vartheta}\frac{2^{\frac{i\vartheta}{2}}\sqrt{\pi}}{\Gamma(\frac{1-i\vartheta}{2})}\\
\end{array}\right)\notag\\
=\left(\begin{array}{cc}
    1-r_{1}(\lambda_{1})r_{2}(\lambda_{1})  &  r_{2}(\lambda_{1})\\
  -r_{1}(\lambda_{1}) &  1\\
\end{array}\right),
\end{gather}
which results in
\begin{align}\label{66}
[M_{1}^{X_{1}}]_{12}=\frac{\sqrt{2\pi}e^{-\frac{3\pi i}{4}+\frac{\pi\vartheta(\lambda_{1}) }{2}}}{ir_{1}(\lambda_{1})\Gamma(i\vartheta(\lambda_{1}))},\quad [M_{1}^{X_{1}}]_{21}=\frac{\sqrt{2\pi}e^{-\frac{\pi i}{4}+\frac{\pi\vartheta(\lambda_{1})}{2}}}{ir_{2}(\lambda_{1})\Gamma(-i\vartheta(\lambda_{1}))}.
\end{align}
Perform similar steps, we have
\begin{align}\label{67}
[M_{1}^{X_{2}}]_{12}=\frac{\sqrt{2\pi}ie^{-\frac{\pi i}{4}+\frac{\pi\vartheta(\lambda_{2}) }{2}}}{r_{1}(\lambda_{2})\Gamma(-i\vartheta(\lambda_{2}))},\quad [M_{1}^{X_{2}}]_{21}=\frac{\sqrt{2\pi}ie^{-\frac{3\pi i}{4}+\frac{\pi\vartheta(\lambda_{2})}{2}}}{r_{2}(\lambda_{2})\Gamma(i\vartheta(\lambda_{2}))},\notag\\
[M_{1}^{X_{3}}]_{12}=\frac{\sqrt{2\pi}e^{-\frac{3\pi i}{4}+\frac{\pi\vartheta(\lambda_{3}) }{2}}}{ir_{1}(\lambda_{3})\Gamma(i\vartheta(\lambda_{3}))},\quad [M_{1}^{X_{3}}]_{21}=\frac{\sqrt{2\pi}e^{-\frac{\pi i}{4}+\frac{\pi\vartheta(\lambda_{3})}{2}}}{ir_{2}(\lambda_{3})\Gamma(-i\vartheta(\lambda_{3}))}.
\end{align}
Substituting them into \eqref{56}, we have
\begin{gather}
q(x,t)=-2i\left[\frac{\sqrt{2\pi}(\delta_{\lambda_{1}}^{0})^{2}e^{-\frac{3\pi i}{4}+\frac{\pi\vartheta(\lambda_{1}) }{2}}}{\sqrt{4t(48\delta\lambda_{1}^{2}-1)}ir_{1}(\lambda_{1})\Gamma(i\vartheta(\lambda_{1}))}\right.\notag\\
\left.+\frac{\sqrt{2\pi}i(\delta_{\lambda_{2}}^{0})^{2}e^{-\frac{\pi i}{4}+\frac{\pi\vartheta(\lambda_{2}) }{2}}}{\sqrt{4t(1-48\delta\lambda_{2}^{2})}r_{1}(\lambda_{2})\Gamma(-i\vartheta(\lambda_{2}))}
+\frac{\sqrt{2\pi}(\delta_{\lambda_{3}}^{0})^{2}e^{-\frac{3\pi i}{4}+\frac{\pi\vartheta(\lambda_{3}) }{2}}}{\sqrt{4t(48\delta\lambda_{3}^{2}-1)}ir_{1}(\lambda_{3})\Gamma(i\vartheta(\lambda_{3}))}\right]+R(\xi,t),\ t\rightarrow \infty.\label{69}
\end{gather}
Finally, we achieve the main result of Theorem 1.1.

\section*{Acknowledgements}
\hspace{0.3cm}
This work was supported by the National Natural Science Foundation of China (No. 12175069 and No. 12235007), Science and Technology Commission of Shanghai Municipality (No. 21JC1402500 and No. 22DZ2229014) and Natural Science Foundation of Shanghai (No. 23ZR1418100).


\begin{thebibliography}{99}
\bibitem{Yang15}
M. J. Ablowitz, Z. H. Musslimani, Integrable nonlocal nonlinear Schr\"{o}dinger equation, Phys. Rev. Lett., 110 (2013) 064105.
\bibitem{Yang16}
M. J. Ablowitz, Z. H. Musslimani, Inverse scattering transform for the integrable nonlocal nonlinear Schr\"{o}dinger equation,
Nonlinearity, 29 (2016) 915-946.
\bibitem{Yang17}
M. J. Ablowitz, Z. H. Musslimani, Integrable nonlocal nonlinear equations, Stud. Appl. Math., 139 (2017) 7-59.

\bibitem{Yang18}
X. Y. Wen, Z. Y. Yan, Y. Q. Yang, Dynamics of higer-order rational solitons for the nonlocal nonlinear Schr\"{o}dinger
equation with the self-induced parity-time-symmetric potential, Chaos, 26 (2016) 063123.

\bibitem{Yang19}
K. Chen, X. Deng, S. Y. Lou, D. J. Zhang, Solutions of nonlocal equations reduced from the AKNS Hierarchy, Stud.
Appl. Math., 141 (2018) 113-141.
\bibitem{Yang20}
B. Yang, Y. Chen, Several reverse-time integrable nonlocal nonlinear equations: Rogue-wave solutions, Chaos, 28 (2018)
053104.
\bibitem{Yang21}
B. F. Feng, X. D. Luo, M. J. Ablowitz, Z. H. Musslimani, General soliton solution to a nonlocal nonlinear Schr\"{o}dinger
equation with zero and nonzero boundary conditions, Nonlinearity, 31 (2018) 5385-5409.
\bibitem{Yang22}
J. L. Ji,  Z. N. Zhu, Soliton solutions of an integrable nonlocal modified Korteweg-de
Vries equation through inverse scattering transform, J. Math. Anal. Appl., 453 (2017) 973-984.
\bibitem{Yang23}
S. Y. Lou, Alice-Bob systems, \^{P}-\^{T}-\^{C} symmetry invariant and symmetry breaking soliton solution, J. Math. Phys., 59
(2018) 083507.
\bibitem{Yang24}
G. Q. Zhang,  Z. Y. Yan, Inverse scattering transforms and soliton solutions of focusing
and defocusing nonlocal mKdV equations with nonzero boundary conditions. Physica
D, 402 (2020) 132170.
\bibitem{Yang25}
M. M. Wang,  Y. Chen, Dynamic behaviors of general N-solitons for the nonlocal generalized nonlinear Schr\"{o}dinger equation. Nonlinear Dyn., 104 (2021) 2621-2638.

\bibitem{He-CTP}
W. Liu,  D. Q. Qiu,  Z. W. Wu, et al. Dynamical Behavior of Solution in Integrable Nonlocal Lakshmanan-Porsezian-Daniel Equation. Commun.  Theor. Phys.,  65(6) (2016) 671.
\bibitem{He-CTP6}
M. Lakshmanan,  K. Porsezian,  M. Daniel, Effect of discreteness on the continuum limit of the Heisenberg spin chain. Phys. Lett. A,  133(9) (1988) 483-488.
\bibitem{He-CTP7}
K. Porsezian, M. Daniel, and M. Lakshmanan, On the integrability aspects of the one-dimensional classical continuum isotropic biquadratic Heisenberg spin chain, J. Math. Phys., 33 (1992) 1807.
\bibitem{He-CTP8}
K. Porsezian, Completely integrable nonlinear Schr\"{o}dinger type equations on moving space curves, Phys. Rev. E, 55 (1997) 3785.

\bibitem{Wave}
X. H. Wu,  Y. T. Gao,  X. Yu, et al. Binary Darboux transformation, solitons, periodic waves and modulation instability for a nonlocal Lakshmanan-Porsezian-Daniel equation. Wave Motion,  114 (2022) 103036.
\bibitem{YangAML}
Y. Yang,  T. Suzuki,  X. Cheng, Darboux transformations and exact solutions for the integrable nonlocal Lakshmanan-Porsezian-Daniel equation. Appl. Math. Lett.,  99 (2020) 105998.
\bibitem{XunTian}
W. K. Xun, S. F. Tian, Inverse scattering transform for the integrable nonlocal Lakshmanan-Porsezian-Daniel equation. arXiv preprint arXiv:2005.04011, 2020.
\bibitem{Peng-Tian9}
C. S. Gardner, J. M. Greene, M. D. Kruskal, R. M. Miura, Method for solving the
Korteweg-de Vries equation. Phys. Rev. Lett., 19 (1967) 1095-1097.
\bibitem{Peng-Tian10}
S. Novikov, S. Manakov, L. Pitaevskii, V. Zakharov, Theory of solitons: the inverse
scattering method. New York and London, Consultants Bureau, 1984.
\bibitem{Peng-Tian15}
S. V. Manakov, Nonlinear Fraunhofer diffraction. Sov. Phys. JETP, 38 (1974) 693-696.
\bibitem{Peng-Tian16}
A. R. Its, Asymptotic behavior of the solutions to the nonlinear Schr\"{o}dinger equation, and isomonodromic deformations of systems of linear differential equations. Sov. Math. Dokl., 24 (1981) 452-456.
\bibitem{Peng-Tian14}
P. A. Deift, X. Zhou, A steepest descent method for oscillatory Riemann-Hilbert
problems. Ann. of Math., 137 (1993) 295-368.

\bibitem{Peng-Tian18}
K. Grunert, G. Teschl, Long-time asymptotics for the Korteweg-de Vries equation via
nonlinear steepest descent. Math. Phys. Anal. Geom., 12 (2009) 287-324.
\bibitem{Peng-Tian19}
P. J. Cheng, S. Venakides, X. Zhou, Long-time asymptotics for the pure radiation
solution of the sine-Gordon equation. Commun. Partial Differential Equations, 24
(1999) 1195-1262.
\bibitem{Peng-Tian20}
A. Boutet de Monvel, A. Kostenko, D. Shepelsky, G. Teschl, Long-time asymptotics
for the Camassa-Holm equation. SIAM J. Math. Anal., 41 (2009), 1559-1588.
\bibitem{Peng-Tian21}
J. Xu, E. G. Fan, Long-time asymptotics for the Fokas-Lenells equation with decaying
initial value problem: without solitons. J. Differ. Equations, 259(3) (2015) 1098-1148.
\bibitem{Peng-Tian22}
B. L. Guo, N. Liu, Y. F. Wang, Long-time asymptotics for the Hirota equation on
the half-line. Nonlinear Anal., 174 (2018) 118-140.
\bibitem{Peng-Tian24}
X. G. Geng,  K. D. Wang,  M. M. Chen, Long-Time Asymptotics for the Spin-1 GrossPitaevskii Equation. Commun. Math. Phys., 382(1) (2021) 585-611.
\bibitem{Peng-Tian25}
S. Y. Chen, Z. Y. Yan, Long-time asymptotics of solutions for the coupled dispersive
AB system with initial value problems. J Math. Anal. Appl., 498(2) (2021) 124966.
\bibitem{Peng-Tian26}
R. Buckingham, S. Venakides, Long-time asymptotics of the nonlinear Schr\"{o}dinger
equation shock problem. Commun. Pure Appl. Math., 60 (2007) 1349-1414.
\bibitem{Peng-Tian27}
Y. L. Yang and E. G. Fan, On the long-time asymptotics of the modified Camassa-Holm equation in space-time solitonic regions. Adv. Math., 402 (2022) 108340.
\bibitem{Peng-Tian28}
Z. Q. Li,  S. F. Tian,  J. J. Yang, On the soliton resolution and the asymptotic stability of N-soliton solution for the Wadati-Konno-Ichikawa equation with finite density initial data in space-time solitonic regions. Adv. Math., 409 (2022) 108639.
\bibitem{Peng-Tian29}
A. Boutet de Monvel, V. P. Kotlyarov, D. Shepelsky, Focusing NLS equation: Longtime dynamics of step-like initial data. Int. Math. Res. Not., (2011) 1613-1653.
\bibitem{Peng-Tian30}
J. Liu, P. A. Perry, and C. Sulem, Long-time behavior of solutions to the derivative nonlinear Schr\"{o}dinger equation for soliton-free initial data. Ann. Inst. Henri Poincare, Anal. Non Lin\'{e}aire, 35 (2018) 217-265.
\bibitem{Peng-Tian31}
Z. Q. Li,  S. F. Tian,  J. J. Yang, Soliton resolution for the Wadati-Konno-Ichikawa equation with weighted Sobolev initial data. Annales Henri Poincar\'{e}. Cham: Springer International Publishing,  23(7) (2022) 2611-2655.
\bibitem{Peng-Tian32}
N. Liu, B. L. Guo, Painlev\'{e}-type asymptotics of an extended modified KdV equation
in transition regions. J Differ. Equations, 280 (2021) 203-235.
\bibitem{Peng-Tian33}
D. S. Wang, B. L. Guo, X. L. Wang, Long-time asymptotics of the focusing Kundu-Eckhaus equation with nonzero boundary conditions. J Differ. Equations, 266(9) (2019) 5209-5253.
\bibitem{Peng-Tian34}
Y. L. Yang, E. G. Fan, Soliton resolution for the short-pulse equation. J Differ. Equations, 280 (2021) 644-689.
\bibitem{JMP-S}
Y. Rybalko, D. Shepelsky, Long-time asymptotics for the integrable nonlocal nonlinear Schr\"{o}dinger equation. J. Math. Phys., 60(3)(2019) 031504.
\bibitem{Peng16-21}
Y. Rybalko,  D. Shepelsky,  Asymptotic stage of modulation instability for the nonlocal nonlinear Schr\"{o}dinger equation. Physica D, 428 (2021) 133060.
\bibitem{Peng16-22}
Y. Rybalko,  D. Shepelsky,  Long-time asymptotics for the nonlocal nonlinear Schr\"{o}dinger equation with step-like initial data. J. Differ. Equations, 270 (2021) 694-724.
\bibitem{Peng16-23}
F. J. He,  E. G. Fan,  J. Xu,  Long-Time asymptotics for the nonlocal MKdV equation. Commun.  Theor. Phys., 71(5) (2019) 475.
\bibitem{Peng-Chen}
W. Q. Peng, Y. Chen, Long-time asymptotics for the reverse space-time nonlocal Hirota equation with decaying initial value problem: Without solitons.	arXiv:2205.10518
\bibitem{JLenells}
J. Lenells, The nonlinear steepest descent method for Riemann-Hilbert problems of low regularity. Indiana Univ. Math., 66 (2017) 1287-1332.
\end{thebibliography}
\end{document}